\documentclass[aps,prl,nofootinbib,twocolumn,superscriptaddress,preprintnumbers,balancelastpage,longbibliography]{revtex4-1}

\usepackage{aas_macros}
\usepackage{titletoc}

\usepackage{placeins}
\usepackage{amsmath,amssymb,mathtools,bm}
\usepackage{graphicx, color, hepunits}
\usepackage[dvipsnames]{xcolor}
\usepackage{cancel}
\usepackage[normalem]{ulem}
\usepackage{float}
\usepackage{filecontents}
\usepackage{multirow}
\usepackage{tikz}
\usetikzlibrary{decorations.pathmorphing}
\usetikzlibrary{arrows}
\usetikzlibrary{shapes.misc}
\usetikzlibrary{positioning}

 \usepackage{hyperref} %Automatically links \label and \ref commands; Always load last
\hypersetup{
    colorlinks=true,       % false: boxed links; true: colored links
    linkcolor=blue,        % color of internal links
    citecolor=blue,        % color of links to bibliography
    filecolor=magenta,     % color of file links
    urlcolor=blue          % color of external links
}
\usepackage[utf8]{inputenc}
\usepackage[english]{babel}
\let\vec\mathbf

\newcommand{\cmt}[1]{}

\newcommand{\es}[2] {\begin{equation} \label{#1} \begin{split} #2 \end{split} \end{equation}}

\newcommand{\new}[1]{\textcolor{black}{#1}}

\begin{document}

\title{
Supernova axions convert to gamma-rays in magnetic fields of progenitor stars}

\author{Claudio Andrea Manzari}
\affiliation{Berkeley Center for Theoretical Physics, University of California, Berkeley, CA 94720, U.S.A.}
\affiliation{Theoretical Physics Group, Lawrence Berkeley National Laboratory, Berkeley, CA 94720, U.S.A.}

\author{Yujin Park}
\affiliation{Berkeley Center for Theoretical Physics, University of California, Berkeley, CA 94720, U.S.A.}
\affiliation{Theoretical Physics Group, Lawrence Berkeley National Laboratory, Berkeley, CA 94720, U.S.A.}

\author{Benjamin R. Safdi}
\affiliation{Berkeley Center for Theoretical Physics, University of California, Berkeley, CA 94720, U.S.A.}
\affiliation{Theoretical Physics Group, Lawrence Berkeley National Laboratory, Berkeley, CA 94720, U.S.A.}

\author{Inbar Savoray}
\affiliation{Berkeley Center for Theoretical Physics, University of California, Berkeley, CA 94720, U.S.A.}
\affiliation{Theoretical Physics Group, Lawrence Berkeley National Laboratory, Berkeley, CA 94720, U.S.A.}

\date{\today}

\begin{abstract}
It has long been established that axions could have been \new{produced within the nascent proto-neutron-star} formed following the type II supernova SN1987A, escaped the star due to their weak interactions, and then converted to gamma-rays in the Galactic magnetic fields; the non-observation of a gamma-ray flash coincident with the neutrino burst 
leads to strong constraints on the axion-photon coupling for axion masses $m_a \lesssim 10^{-10}$ eV. In this work we use SN1987A to \new{constrain higher mass axions,} all the way to $m_a \sim 10^{-3}$ eV, \new{by accounting for axion production from the Primakoff process, nucleon bremsstrahlung, and pion conversion along with axion-photon conversion on the still-intact magnetic fields of the progenitor star.}  Moreover, we show that gamma-ray observations of the next Galactic supernova, leveraging the magnetic fields of the progenitor star, could detect quantum chromodynamics axions for masses above roughly $50$ $\mu$eV, depending on the supernova. We propose a new full-sky gamma-ray satellite constellation that we call the GALactic AXion Instrument for Supernova (GALAXIS) to search for such future signals along with related signals from extragalactic neutron star mergers. 
\end{abstract}
\maketitle

Supernova (SN) 1987A (SN1987A) was a type II SN that exploded in February 1987, producing roughly two dozen neutrino events that were detected at the Kamiokande II, IMB, and Baksan neutrino detectors over a time interval of around 10 s~\cite{Kamiokande-II:1987idp,Bionta:1987qt,Alekseev:1987ej}.  The SN took place in the Large Magellanic Cloud at a distance of approximately 51.4 kpc from Earth.  
 SN1987A 
 provides some of the most stringent and well-established constraints on a class of hypothetical ultra-light pseudo-scalar particles known as axions~\cite{Raffelt:1996wa,Raffelt:1990yz,Brockway:1996yr,Grifols:1996id,Payez:2014xsa,Hoof:2022xbe,Caputo:2024oqc}.  These constraints have been made all the more robust recently by the tentative discovery of the NS formed after SN1987A, helping establish that the SN formed a neutron star (NS) and not a black hole~\cite{Page:2020gsx,Fransson:2024csf}.
 In this work we point out for the first time a novel constraint from SN1987A that has promising implications for future SN; axions produced within the proto-NS (PNS) can convert to observable gamma-rays in the stellar magnetic field of the progenitor star.
 
 \begin{figure}[!tp]
    \centering
\includegraphics[width=1.0\columnwidth]{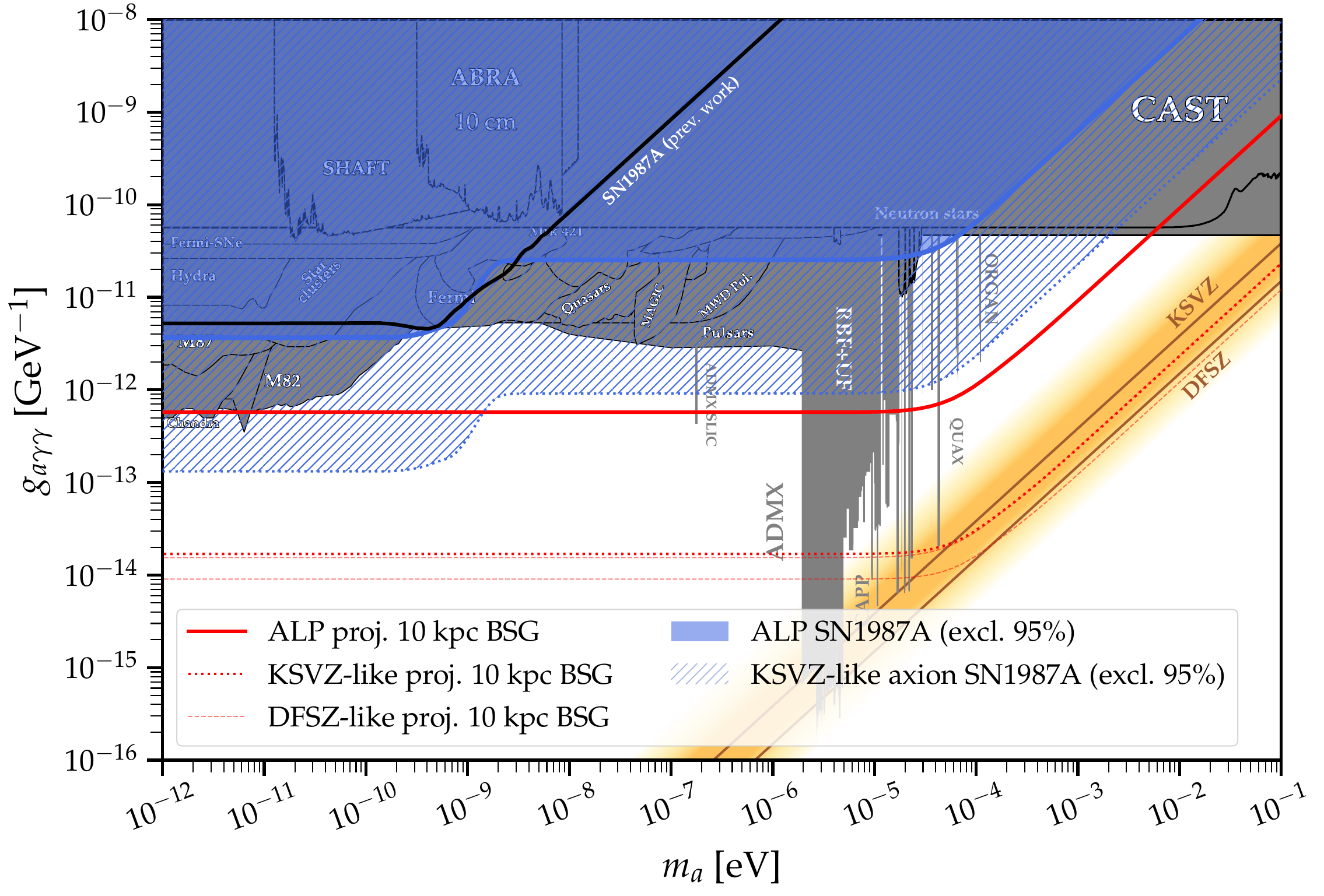}
\caption{Existing constraints (notably~\cite{Reynolds:2019uqt,Dessert:2020lil,Reynes:2021bpe,Noordhuis:2022ljw,Dessert:2022yqq,Dolan:2022kul,Davies:2022wvj,Ning:2024eky} and~\cite{ciaran_o_hare_2020_3932430,Workman:2022ynf} for reviews) on the axion-photon coupling $g_{a\gamma\gamma}$ as a function of the axion mass $m_a$ are shaded in grey, with the previously-leading constraint from the non-observation of axion-induced gamma-rays from SN1987A highlighted~\cite{Payez:2014xsa,Hoof:2022xbe}.  We 
point out in this work that the axions could convert to gamma-rays in the stellar magnetic field of the progenitor star, extending the upper limit on $g_{a\gamma\gamma}$ to higher masses as indicated in shaded blue.  We take the surface field strength of the progenitor to be 100 G to be conservative ($\sim$1 kG is favored).  Note that the KSVZ-like axion model assumes the couplings to photons and hadrons are related as in the KSVZ QCD-axion model; \new{the axion-like particle (ALP) model assumes loop-induced hadronic couplings (see text).} 
The non-observation of gamma-rays from the next Galactic SN (assumed to be at $d = 10$ kpc) with the proposed GALAXIS full-sky gamma-ray telescope network (modeled as being equivalent to the on-axis Fermi-LAT instrument response with full-sky coverage) could cover vast regions of QCD axion parameter space (red), depending on the properties of the progenitor star (BSG shown here, assuming a typical 1 kG surface field strength) and the axion.    
}
\label{fig:SN_limit_projs}
\end{figure}

 Axions 
 may address a number of outstanding problems in nature such as the strong-{\it CP} problem~\cite{Peccei:1977hh,Peccei:1977ur,Weinberg:1977ma,Wilczek:1977pj} ({\it i.e.}, the lack of a neutron electric dipole moment) and the measured dark matter abundance in the Universe~\cite{Preskill:1982cy,Abbott:1982af,Dine:1982ah}. Moreover, axions are now understood to arise generically in
 string theory compactifications~\cite{Svrcek:2006yi, Arvanitaki:2009fg,Cicoli:2012sz,Demirtas:2018akl,Halverson:2019cmy,Demirtas:2021gsq}. String theory motivates the picture of the `axiverse,' where the quantum chromodynamics (QCD) axion that solves the strong-{\it CP} problem is accompanied by a number of axion-like particles, which interact through higher dimensional operators with the rest of the Standard Model but not with QCD. 
The QCD axions receive a mass contribution from QCD  
of the order $m_a^{\rm QCD} \approx 5.70 \, \, {\rm \mu eV} ( 10^{12} \, \, {\rm GeV} / f_a )$, with $f_a$ the axion decay constant.   
The axion field $a$ has an interaction with photons ${\mathcal L} = g_{a\gamma\gamma} a {\bf E} \cdot {\bf B}$, with ${\bf E}$ (${\bf B}$) the electric (magnetic) field, 
that is parameterized by the coupling constant \mbox{$g_{a\gamma\gamma} \equiv C_{a\gamma\gamma} \alpha_{\rm EM} / (2 \pi f_a)$}, with $\alpha_{\rm EM}$ the fine-structure constant and $C_{a\gamma\gamma}$ a coefficient of order unity that depends on the ultraviolet (UV) completion. 
For the QCD axion we thus expect $g_{a \gamma\gamma} \propto m_a$, as illustrated by the gold band in Fig.~\ref{fig:SN_limit_projs}; 
axion-like particles 
are motivated throughout the $g_{a\gamma\gamma}$-$m_a$ plane.   
 
 There are two classes of well-established constraints on the interaction strengths of light axions ($m_a \lesssim {\rm eV}$) with the Standard Model from SN1987A: (i) axion production in the PNS core can modify the thermal evolution of the PNS, modifying the predicted luminosity evolution of neutrinos~\cite{Raffelt:1987yt,1987PhLB..193..525E,Turner:1987by,1988PhLB..203..188M,1996bboe.book..188M,Brinkmann:1988vi,Burrows:1988ah,Raffelt:1990yz,Janka:1995ir,Keil:1996ju,Fischer:2016cyd,Carenza:2019pxu,Lella:2023bfb, Manzari:2023gkt}; and (ii) 
 ultralight axions that escape the PNS core could later convert to gamma-rays in Galactic magnetic fields~\cite{Raffelt:1996wa,Brockway:1996yr,Grifols:1996id,Payez:2014xsa,Hoof:2022xbe,Caputo:2024oqc} (see also~\cite{Meyer:2016wrm,Calore:2023srn} for prospects for future SN). The latter probe is supported by the non-observations of gamma rays coincident with the neutrino burst by the Solar Maximum Mission (SMM)~\cite{1980SoPh...65...15F}, which happened to be looking in the direction of SN1987A when the explosion took place.  In this work we propose a third probe of axions from SN1987A, future SN, and even NS-NS mergers that relies on axion-photon conversion in the stellar magnetic fields of the progenitor star. This third probe allows us to test an intermediate axion mass range, extending up to $m_a \sim 10^{-1}$ eV, as indicated in Fig.~\ref{fig:SN_limit_projs}.

In addition to considering SN1987A, we perform projections for the next Galactic SN and demonstrate that if an instrument such as the Fermi Large Area Telescope (LAT) were to observe such an event the non-observation of coincident gamma-rays could rule out or detect QCD axions above roughly $50$  $\mu$eV, depending on the precise properties of the SN, by accounting for axion-to-photon conversion on the progenitor's magnetic fields. On the other hand, its limited field of view (FOV) means that Fermi-LAT only has around a one in five chance of fortuitously looking at the right place at the right time to catch the next Galactic SN. Given that the Galactic SN rate is around one per hundred years, we thus find ourselves unprepared to take advantage of this rare event for axion physics. To address this shortfall we propose a network of space-based gamma-ray telescopes in the 100's of MeV range to search for gamma-ray flashes from Galactic SN and similar nearby extragalactic events, such as NS-NS mergers; we refer to this network as the GALactic AXion Instrument for Supernova (GALAXIS).  

We make a number of improvements in modeling axion-induced
gamma-ray signals from PNSs.
For axion-like particles that couple only to electroweak gauge bosons in the UV, we show that their infrared (IR) renormalization-group induced couplings to quarks typically dominate the axion production rate within the PNS. Two classes of production mechanisms are important for this result: (i) axion production from nucleon bremsstrahlung, and (ii) axion production from pion conversion. (See~\cite{Carenza:2020cis,Fischer:2021jfm,Choi:2021ign,Vonk:2022tho,Ho:2022oaw} for previous discussions of pion-induced axions in SN.)  The QCD axion has tree-level couplings to nucleons and pions, and accounting for these interactions is crucial in projecting the sensitivity of proposed future SN observations to QCD axions.  Additionally, we make use of a suite of cutting-edge SN simulations~\cite{Bollig:2020xdr} that are spherically symmetric but include PNS convection, muons and muon neutrinos, general relativity, and neutrino transport~\cite{Janka:2012wk, Bollig:2017lki}.

{\bf Axion luminosity from a PNS.---}
The effective field theories (EFTs) for the QCD axion and for axion-like particles contain the interactions 
\mbox{${\mathcal L} \supset {g_{aqq} \over 2 m_q} (\partial^\mu a) \bar q \gamma_\mu \gamma_5 q$}, where 
$g_{aqq} = C_{aqq} m_q / f_a$, with $C_{aqq}$ a UV-dependent coefficient and $m_q$ the quark masses for quark fields $q$.  There are additional interactions involving leptons, but they are not relevant for this work. 
The QCD axion additionally has the coupling ${\mathcal L} \supset {g^2 \over 32 \pi^2 f_a} a G_{\mu \nu}^a \tilde{G}^{a\,\mu \nu}$, which involves the QCD field strength $G_{\mu\nu}^a$ and the strong-coupling-constant $g$.

Below the scale of the QCD phase transition it is more instructive to talk about the axion couplings to hadrons than to quarks. Moreover, the axion and $\pi^0$ undergo a mass mixing for the QCD axion, which provides an IR contribution to $C_{a\gamma\gamma}$. The axion-nucleon couplings are of the same form as the axion-quark couplings but with coefficients $C_{app}$ and $C_{ann}$ for the proton and neutron, respectively. The axion-pion-nucleon interaction may be computed in heavy baryon chiral perturbation theory~\cite{Chang:1993gm,Vonk:2021sit} and reads 
\es{}{
\mathcal{L}_{a\pi N} = i \frac{\partial_{\mu} a}{2 f_a} C_{a\pi N} (\pi^+ \bar{p}\gamma^{\mu} n - \pi^{-}\bar{n}\gamma^{\mu}p)\,,
}
with $C_{a\pi N} = (C_{app}-C_{ann})/{\sqrt{2}g_A}$, where $g_A\approx 1.28$ is the axial-vector coupling constant.

The QCD axion necessarily has tree-level couplings to hadrons because of the axion-gluon coupling.  In KSVZ-type models~\cite{Kim:1979if,Shifman:1979if}, where the axion does not couple at tree-level to fermions in the UV, $C_{app} \approx -0.47$, $C_{ann} \approx -0.02$, and $C_{a \pi N} \approx -0.27$. 
 In DFSZ-type models~\cite{Dine:1981rt,Zhitnitsky:1980tq} where there are UV couplings of the axion to fermions, with $\tan \beta$ the ratio of up-type to down-type vacuum expectation values of the two Higgs doublets in those models, the axion-matter couplings can be further enhanced; see the Supplementary Material (SM).  

Axion-like particles may or may not have UV contributions to the axion-quark and hence axion-nucleon couplings. On the other hand, even if all $C_{aqq} = 0$ at the PQ scale $f_a$, these operators are generated under the renormalization group (RG) flow, leading to non-zero values for $C_{app}$, $C_{ann}$, and $C_{a \pi N}$ in the IR~\cite{Bauer:2021mvw}.  The precise IR values for these loop-induced coefficients depends on how the axion couples to $SU(2)_L$ and $U(1)_Y$ gauge bosons; as we discuss further in the SM, a generic expectation for the loop-induced coefficients is  \mbox{$C_{app} / C_{a\gamma\gamma} \approx C_{ann} / C_{a\gamma\gamma} \approx 10^{-4}$} and \mbox{$C_{a \pi N} / C_{a\gamma\gamma} \approx 10^{-5}$}.  We adopt these choices to be conservative, since this assumes no additional UV contributions, when discussing axion-like particle models. \new{Note that the limit from white dwarfs in Ref.~\cite{Dessert:2021bkv} in Fig.~\ref{fig:SN_limit_projs} also used loop-induced couplings -- in that case to electrons; on the other hand, the other upper limits shown in Fig.~\ref{fig:SN_limit_projs} are not enhanced or otherwise affected by assuming loop-induced couplings to matter.  }

 Hot PNSs have thermal populations of photons, nucleons, and pions. 
 These populations may produce axions through the Primakoff process (for photons), bremsstrahlung (for nucleons), and through pion-to-axion conversion off of nucleons, either through the four-point interaction or through intermediate nucleon or $\Delta$ resonances~\cite{Raffelt:1985nk, Carenza:2019pxu, Ho:2022oaw}.

We improve the calculation of the axion luminosity relative to previous works on gamma-ray signals from SN1987A ({\it e.g.},~\cite{Payez:2014xsa, Fischer:2021jfm}) by making use of more modern SN simulations. In particular, we use the SN simulations presented in Ref.~\cite{Bollig:2020xdr}, whose radial profiles are accessed through the Garching Core-Collapse Supernova archive~\cite{GCCSN}. (See also the recent SN simulations in~\cite{Fiorillo:2023frv}.) These are spherically symmetrical (1D) models that include PNS convection~\cite{Mirizzi:2015eza}, the presence of muons and muon-neutrinos, general relativity, and neutrino transport~\cite{Rampp:2002bq,Janka:2012wk, Bollig:2017lki}. 

To assess the impact of the astrophysical uncertainties related to the mass of NS1987A formed by SN1987A, we consider three different simulations: SFHo-18.6, SFHo-18.8 and SFHo-20.0. Model SFHo-18.6, which is our fiducial model,  assumes an 18.6 ${\rm M}_\odot$  progenitor and has a NS mass of 1.553 ${\rm M}_\odot$, well within the range expected for NS 1987A ({\it e.g.},~\cite{Page:2020gsx}). Model SFHo-18.8  assumes an 18.8 ${\rm M}_\odot$  progenitor, and the remnant NS mass is 1.351 ${\rm M}_\odot$, at the lower edge of the expected range, while in model SFHo-20.0 the progenitor star has a mass of 20 ${\rm M}_\odot$ and the NS mass is 1.947 ${\rm M}_\odot$, near the upper edge of the expected range. 
The SFHo equation of state (EOS) that is implemented in these simulations is fully compatible with all current constraints from nuclear theory and experiment~\cite{Fischer:2013eka,Oertel:2016bki,Fischer:2017zcr} and astrophysics, including pulsar mass measurements~\cite{Demorest:2010bx,Antoniadis:2013pzd,NANOGrav:2019jur} and the radius constraints deduced from gravitational-wave and Neutron Star Interior Composition Explorer (NICER) measurements~\cite{LIGOScientific:2018cki, Bauswein:2017vtn,Essick:2020flb}.

The simulations cover the first $\sim$$10\; {\rm s}$ after bounce, with the explosion triggered at $t \sim 0.16 \;{\rm s}$. The data are provided in intervals of 0.025 s for $0\; {\rm s}< {\rm t} < 0.5\; {\rm s}$, in intervals of 0.25 s for $0.5\; {\rm s}< {\rm t} < 3\; {\rm s}$, in intervals of 0.5 s for $3\; {\rm s}< {\rm t} < 6\; {\rm s}$, and in intervals of 1 s until the end of the simulation.  
The radially-dependent temperature peaks around 40 MeV at $\sim$1 s after the explosion and maintains a temperature $\gtrsim$5 MeV until 10 s after. 

\begin{figure}[!t]
    \centering
\includegraphics[width=0.95\columnwidth]{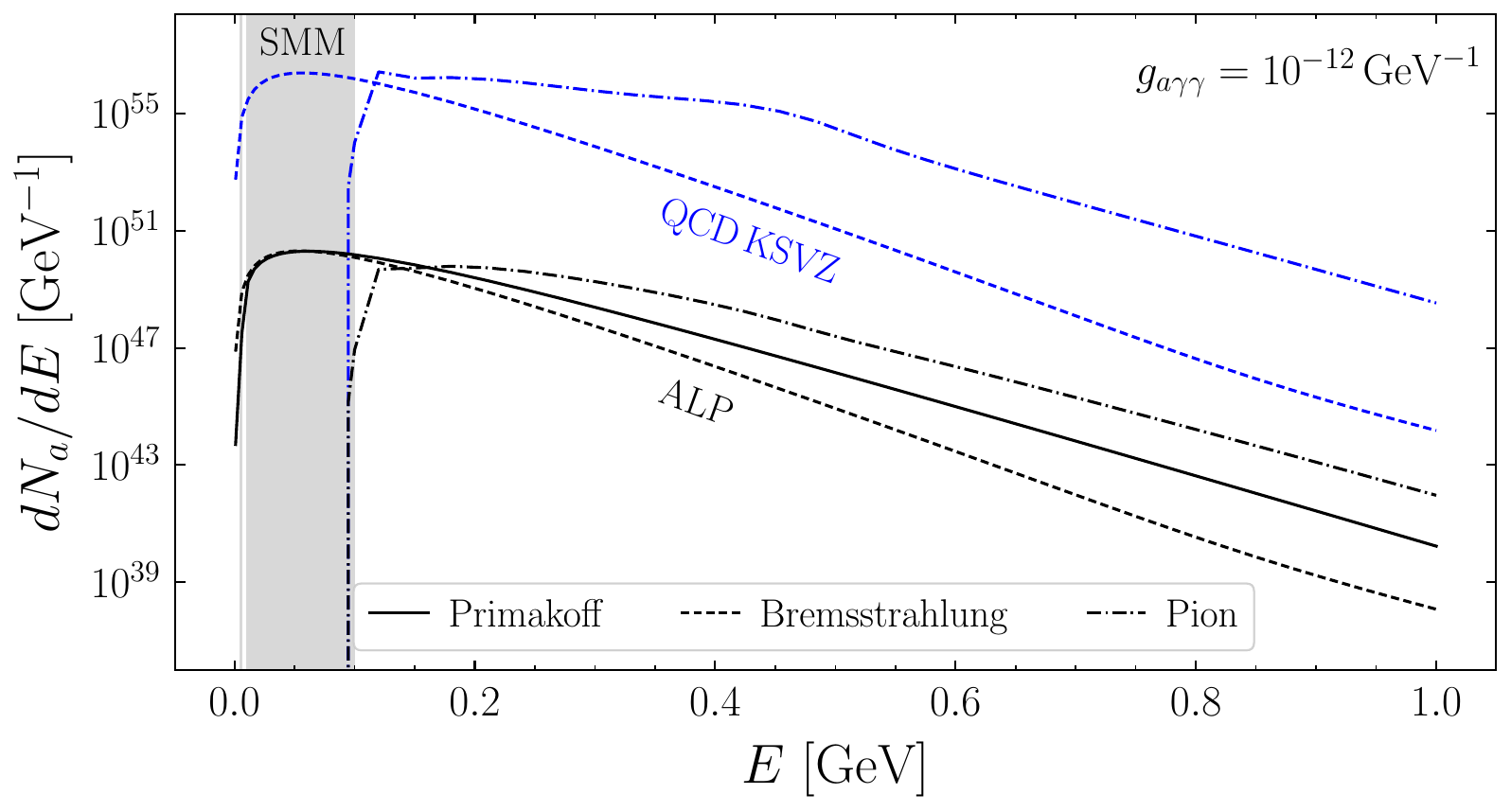}
    \caption{The differential axion spectra integrated over the first 10 s after the SN for our fiducial SN1987A simulation SFHo-18.6~\cite{Bollig:2020xdr}, corresponding to the formation of a 1.553 $M_\odot$ NS.  We separate the spectra into contributions from the Primakoff production, bremsstrahlung from nucleons, and processes involving pions. 
    The ALP curves assume no UV contributions to the axion-quark couplings, with the couplings generated in the IR under the RG flow, while the curve labeled `QCD KSVZ' uses the relations $C_{a nn} / C_{a\gamma\gamma} \simeq 0.01$, $C_{a pp} / C_{a\gamma\gamma} \simeq 0.24$ and $C_{a\pi N} / C_{a\gamma\gamma} \simeq 0.13$ appropriate for a KSVZ-type QCD axion. \new{By construction the Primakoff curve is common to both models.}  }
    \label{fig:luminosities}
\end{figure}

We compute the axion luminosities in each time slice of the simulation using the radial profiles of the temperature and the chemical potentials.
In Fig.~\ref{fig:luminosities} we illustrate the differential
axion spectra $dN_a / dE$ integrated over the 10 s simulation (our fiducial model) for two different theory assumptions for the axion. Both cases have $g_{a\gamma\gamma} = 10^{-12}$ GeV$^{-1}$, but that labeled `ALP' has no tree-level coupling to quarks and only the loop-induced couplings described previously.  The second case, labeled `QCD KSVZ', has $C_{a nn} / C_{a\gamma\gamma} \simeq 0.01$, $C_{a pp} / C_{a\gamma\gamma} \simeq 0.24$ and $C_{a\pi N} / C_{a\gamma\gamma} \simeq 0.13$, which are the ratios expected in the KSVZ QCD axion model. (See also the `light' QCD axion models proposed in~\cite{Hook:2018jle,DiLuzio:2021pxd}.)
For each scenario we show the contributions to the luminosity from Primakoff production, nucleon bremsstrahlung involving nucleons only, and processes involving pions. Interestingly, even in the axion-like particle scenario with no tree-level fermion couplings the contribution to the luminosity from hadrons is comparable to the Primakoff production. 
We also indicate the energy range of the SMM telescope that observed SN1987A;  the majority of the pion-induced emission is outside of SMM's energy range. 

{\bf Axion-photon conversion.---}
We consider, for the first time, 
the conversion of axions-to-photons on the stellar magnetic fields surrounding the progenitor star for the SN. 
First, it is instructive to make a rough estimate of the Galactic versus stellar conversion probabilities, with the low-axion-mass approximation $P_{a\to\gamma} \sim g_{a\gamma\gamma}^2 B^2 L^2$, with $B$ the astrophysical magnetic field strength and $L$ the length of the magnetic field domain. Typical values for Galactic magnetic fields are $B \sim \mu$G and $L \sim 1$ kpc, yielding $P_{a \to \gamma} \sim 10^{-5} (g_{a \gamma \gamma} / 10^{-12} \, \, {\rm GeV})^2$.  On the other hand, the progenitor of the SN1987A was a blue supergiant (BSG), with a surface magnetic field strength $B_0 \sim {\rm kG}$~\cite{Orlando:2018lbj} and a radius $r_0 \approx 45 \pm 15$ $R_\odot$~\cite{1988ApJ...330..218W}. (We fix $r_0 = 45$ $R_\odot$ as this is a subdominant source of uncertainty relative to the surface magnetic field strength.)  Given that $\mu {\rm G} \times {\rm kpc} \sim {\rm kG} \times (45 R_\odot)$, we estimate that the axion-to-photon conversion probability on the stellar magnetic fields should be comparable to that on the Galactic fields.  On the other hand, the estimates above are only valid in the low mass limit; in particular, they are valid when $m_a^2 / (2 E) \times L \ll 1$, where $E$ is the energy of the axion.  Taking $E \sim 100$ MeV, we thus estimate that the axion-conversion probability becomes degraded for $m_a \gtrsim 2 \times 10^{-11}$ eV ($m_a \gtrsim 5 \times 10^{-5}$ eV) for conversion on the Galactic (stellar) magnetic fields.   

Core-collapse supernova form PNS 
when the collapsing core reaches nuclear densities; the formation of the PNS causes the in-falling matter to bounce outwards, forming a rapidly expanding shock-wave that blows apart the star. The outward propagating shock-wave travels slower than the speed of light. In contrast, the axions
propagate outwards faster, nearly at the speed of light.  Thus,
the axions leave the star well ahead of the shock-wave. They encounter the still-pristine magnetic fields of the progenitor star because the change in the magnetic field induced by the bounce propagates relatively slowly out from the stellar core at the Alfv\'en velocity (see, {\it e.g.},~\cite{Suzuki:2007ej}).

There was no direct measurement of the magnetic field strength of the SN1987A precursor star Sk -69 202, \new{but there is indirect evidence from combining radio and X-ray data in the decades following the SN with models for the expanding SN remnant that the precursor star had a surface field strength $B_0 \sim 3$ kG~\cite{Orlando:2018lbj,Petruk:2022urq}.}  This field strength is in line with the $\sim$kG level magnetic field strengths expected for BSGs~\cite{2015A&A...584A..54P}, especially considering that Sk -69 202 likely formed from a merger of two smaller stars~\cite{2012Sci...337..444S,2007Sci...315.1103M}. BSGs like Sk -69 202 have surface field strengths in the range $\sim$100 G to 10 kG~\cite{2009ARA&A..47..333D}, and below we use this range of field strengths when bracketing the uncertainties in the axion-induced gamma-ray signal.  \new{In partricular, we estimate from population synthesis data that less than $\sim$3\% of BSGs like SK -69 202 have dipole field strengths less than 100 G~\cite{2015A&A...574A..20F}, such that 100 G may be considered a robust lower bound on the dipole field strength.} 

We model the magnetic field of the progenitor star as a dipole field, in which case $B$ falls as $1/r^3$ away from the stellar surface. On the other hand, we note that this is a conservative choice, as the rotation of the progenitor star and its stellar wind may have led to a Parker Spiral type field~\cite{1958ApJ...128..664P,Orlando:2018lbj}, as in the case of the Sun, for which $B$ falls more slowly, with $1/r$ and $1/r^2$ components, away from the surface. 
We assume for simplicity that the axions travel radially outwards at the mid-plane, such that at every point exterior to the star the magnetic field is perpendicular to the axion trajectory.  The axion-photon mixing equations are described in detail in the SM, including the non-linear Euler-Heisenberg (EH) term in the effective Lagrangian for electromagnetism (see, {\it e.g.},~\cite{Safdi:2022xkm}), which reduces the conversion probabilities given the large axion energies and high field strengths.  Note that we neglect the effects of the photon plasma frequency in the medium exterior to the star, since for this to be important the free electron density would need to exceed $n_e \sim 10^{10}$ cm$^{-3}$, which is not expected.

{\bf SMM data analysis from SN1987A.---}
We 
compute the mass-dependent upper limits on $g_{a\gamma\gamma}$ from the non-observation of excess gamma-rays from SN1987A with the SMM. We use the SMM data and instrument response approximations presented in~\cite{Hoof:2022xbe}.  (See the SM.)  We find no evidence for axions, consistent with previous works, with the 95\% upper limits illustrated in Fig.~\ref{fig:SN_limit_projs}. The limit shaded in blue labeled `ALP SN1987A' accounts both for the conversion in the Galactic magnetic field, with the fields modeled using the updated Galactic model~\cite{Unger:2023lob} (for each mass and energy point we use the lowest conversion probability among all models described in Ref.~\cite{Unger:2023lob}), and for axion-to-photon conversion in the stellar magnetic field of the progenitor (dominating the sensitivity for $m_a$ above $\sim$$10^{-9}$ eV).  
We account for 
Primakoff production and hadronic processes, with our fiducial loop-level couplings to hadrons.  Only accounting for Primakoff emission weakens the limit at low $m_a$ from  $|g_{a \gamma \gamma}| \lesssim 2.6 \times 10^{-12}$ GeV$^{-1}$ to $|g_{a \gamma \gamma}| \lesssim 3.1 \times 10^{-12}$ GeV$^{-1}$.  On the other hand, changing 
to the older Galactic magnetic field model in Ref.~\cite{Jansson:2012pc}, matching that used in previous works~\cite{Payez:2014xsa,Hoof:2022xbe}, weakens the low-mass axion limit to $|g_{a \gamma \gamma}| \lesssim 3.4 \times 10^{-12}$ GeV$^{-1}$.  Using the magnetic field model in~\cite{Jansson:2012pc} and only accounting for Primakoff emission, as in Refs.~\cite{Payez:2014xsa,Hoof:2022xbe}, we find a nearly identical upper limit to that in~\cite{Hoof:2022xbe} at $m_a = 0$ eV ($<$ 10\% difference), suggesting that the differences in SN simulations are sub-leading. 
 
Our upper limits in Fig.~\ref{fig:SN_limit_projs} take a stellar surface field strength of 100 G to be conservative, even though higher field strengths are favored.  Our axion-like particle limit (labeled `ALP SN1987A') excludes new parameter space for $m_a \gtrsim 10$ $\mu$eV.
The upper limit  labeled `KSVZ-like axion SN1987A' assumes that the ratios of $C_{app} / C_{a\gamma\gamma}$, $C_{ann} / C_{a\gamma\gamma}$, and $C_{a\pi N} / C_{a\gamma\gamma}$ are as expected in the KSVZ QCD axion model. 
This upper limit is around an order of magnitude away in terms of $g_{a\gamma\gamma}$ from probing the KSVZ QCD axion model for $m_a \gtrsim 10^{-4}$ eV, strongly motivating future observations with increased sensitivity.

Our axion-like particle upper limit in Fig.~\ref{fig:SN_limit_projs} excludes much of the parameter space that will be probed by the ALPS II light-shining-through-walls experiment~\cite{Bahre:2013ywa,Oceano:2024maq,Ringwald:2024uds}; taking a more realistic but less conservative surface magnetic fields strength of $B_0 = 1$ kG, we exclude the full parameter space to be probed by ALPS II (see the SM).

Note that our results are strictly speaking only valid for, roughly, $|g_{a\gamma\gamma}| \lesssim 10^{-8}$ GeV$^{-1}$ ($m_a \lesssim 10^{-2}$ eV) for the axion-like particle model (the KSVZ QCD axion), as for larger couplings we estimate that the axion luminosity 1 s post-bounce exceeds the neutrino luminosity.  The axion model at larger couplings is disfavored~\cite{Raffelt:1996wa,Caputo:2024oqc}, and modeling this scenario would require including the back-reaction of the axion emission in the SN simulations.

{\bf GALAXIS: Galactic Axion Instrument for Supernova.---} 
If a Galactic SN went off today, we estimate \new{using the \texttt{Fermitools}\footnote{\url{https://fermi.gsfc.nasa.gov/ssc/data/analysis/documentation/}}} that the chance  Fermi-LAT would be looking at the correct place at the correct time to catch the $\sim$10 s axion-induced burst is only around $\sim$20\%, accounting for the finite FOV of the instrument and down-time during its orbit. 
On the other hand, if the next SN went off directly above Fermi (at its zenith), the estimated 95\% upper limits on $g_{a\gamma\gamma}$ we would be able to obtain are illustrated in Fig.~\ref{fig:SN_limit_projs}. 
We use the \texttt{Fermitools} to obtain the instrument response with the \texttt{P8R3\_TRANSIENT020\_V2} event class; we estimate $\sim$0 background events over the $\sim$10 s duration of the SN.  The effective area at zenith at $E = 200$ MeV is $\sim$$0.72$ m$^2$.
We illustrate the expected 95\% upper limits under the null hypothesis for the axion-like particle scenario, the KSVZ-like axion, and a DFSZ-like scenario, scanning over $\tan \beta$. (We show the strongest and weakest limits across the range of $\tan \beta$; see the SM for details.)  We make these projections using our fiducial SFHo-18.6 SN simulation, and we assume a distance of 10 kpc to the next Galactic SN. We only account for axion-photon conversion on the stellar magnetic fields of the progenitor star, assuming a 1 kG surface magnetic field for a BSG SN that is otherwise the same as SN1987A. (The axions could also convert to photons on the Galactic magnetic field, enhancing the low-mass sensitivity.) In the SM we discusses red supergiant (RSG) SN, which are more prevalent and as we show have comparable sensitivity.

\begin{figure}[!t]
    \centering
\includegraphics[width=1.0\columnwidth]{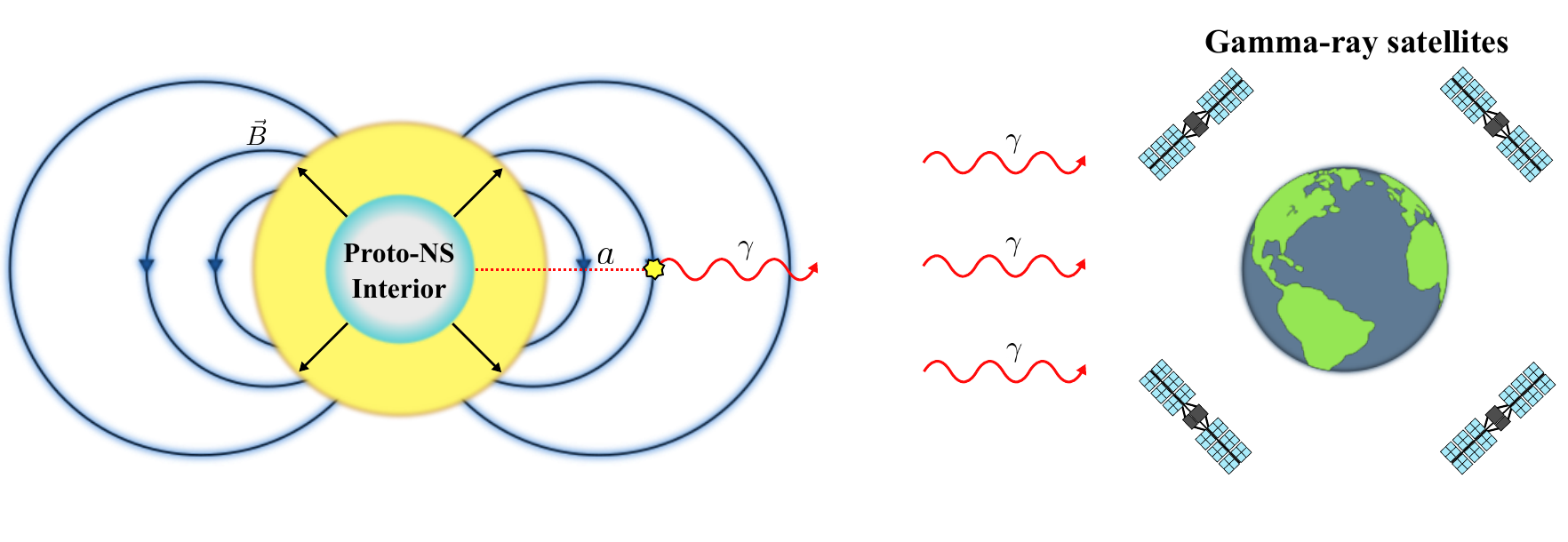}
    \caption{The GALAXIS gamma-ray satellite constellation proposed in this work to search for axion-induced gamma-ray signatures from core-collapse SN and NS-NS mergers. The axions are generated in the hot PNS cores and then convert to gamma-rays on the stellar magnetic fields of the progenitor stars. Such an instrument with a Fermi-LAT-level effective area could potentially probe QCD axions for any $m_a \gtrsim 50$ $\mu$eV, depending on the properties of the event.}
    \label{fig:sat}
\end{figure}

Without new instrumentation the opportunity to probe QCD axions using gamma-ray observations of the next Galactic SN will almost certainly be lost, since the event will likely have no advanced warning (but see~\cite{Ge:2020zww}) and not be within the FOV of the Fermi-LAT.  The proposed Advanced Particle-astrophysics Telescope (APT)~\cite{Alnussirat:2021tlo,APT:2021lhj} may have an increased FOV relative to the Fermi-LAT, though it will likely also not be $4 \pi$. We thus propose a full-sky gamma-ray telescope network, which we call the GALactic AXion Instrument for Supernova (GALAXIS) (see Fig.~\ref{fig:sat}). 

The idea behind GALAXIS is to establish a full-sky constellation of gamma-ray satellites to provide continuous $4 \pi$ coverage of the gamma-ray sky between $\sim$100 MeV and $\sim$1 GeV. (See also the recent work~\cite{Lella:2024hfk} that made a related proposal.)  The network would consist of multiple ({\it e.g.}, $\sim$5 or more) gamma-ray telescopes on different orbital trajectories, such that any future SN would be in view of at least one telescope in the network.  Such an instrument would complement the multiple gamma-ray telescope constellations in planning stages at energies below $\sim$10 MeV (see~\cite{Bloser:2022pnu} and references therein).  We leave a full technical investigation to future work. 
 In Fig.~\ref{fig:SN_limit_projs} we simply assume for the projections that the GALAXIS instrument response is identical to that of the on-axis Fermi-LAT (see the SM). 
 The main improvements with the future projections relative to the SN1987A constraints come from the distance to the SN, the large effective area and improved background rate of GALAXIS ({\it i.e.}, Fermi-LAT) relative to the SMM, and the inclusion of higher-energy photons 
above $\sim$100 MeV that allow for probing pion-induced axions. GALAXIS may reach sensitivity to the QCD axion, making it competitive with upcoming efforts to target QCD axions such as IAXO~\cite{IAXO:2019mpb}, MADMAX~\cite{Beurthey:2020yuq}, and ALPHA~\cite{Millar:2022peq}.

{\bf Discussion.---}
In this Letter we focus on axion-induced gamma-ray signals from nearby PNS formed after core-collapse SN due to axion-photon conversion in the stellar magnetic fields of the progenitor stars.  However, there are a number of related axion-induced gamma-ray signals that may proceed similarly and be detectable with the proposed GALAXIS gamma-ray observatory.  For example, 
 in cases where the compact remnant of the core-collapse SN is a black hole (as suggested could be the case for SN1987A in~\cite{Bar:2019ifz}, though this is now disfavored~\cite{Page:2020gsx,Fransson:2024csf}), a hot, massive PNS remnant forms prior to collapse.  It would be interesting to study the axion-induced gamma-ray signal from such a short-lived remnant with dedicated simulations.  Similarly, NS-NS mergers can lead to stable NSs or hypermassive remnants that collapse to black holes; in either case, exceedingly hot PNSs form within the first tens of ms, with temperatures that can exceed those in core-collapse SN. As we show in the SM, nearby NS-NS mergers (within $\sim$50 Mpc of Earth) are promising targets for gamma-ray axion searches. (See also~\cite{Dietrich:2019shr,Harris:2020qim,Dev:2023hax}.) Given the compact sizes of the NSs, these objects can potentially probe higher axion masses and may even reach QCD axion sensitivity near $\sim$1 meV (see the SM).  NS-NS mergers, along with SN within the local group and nearby galaxy clusters, can be expected on a near yearly basis, meaning that the proposed GALAXIS instrument would have frequent opportunities for axion science.   

\let\oldaddcontentsline\addcontentsline% Store \addcontentsline
\renewcommand{\addcontentsline}[3]{}% Make \addcontentsline a no-op

\section*{Acknowledgements}

{\it
We thank L. Bildsten, A. S. Brun, J. Buckley, A. Caputo, F. Calore, P. Carenza, E. Charles, A. Chen, C. Dessert, A. Filippenko, D. F. G. Fiorillo, J. Foster, S. Hoof, A. Lella, A. Mirizzi, E. Most, O. Ning, A. Prabhu, G. Raffelt, A. Ringwald, A. Spitkovsky, J.  Tomsick, and S. Witte for helpful discussions.
B.R.S and Y.P. were supported in part by the DOE Early Career Grant DESC0019225.  C.A.M. and I.S. were supported by the Office of High Energy Physics of the U.S. Department of Energy under contract DE-AC02-05CH11231
This research used resources of the National Energy Research Scientific Computing Center (NERSC), a U.S. Department of Energy Office of Science User Facility located at Lawrence Berkeley National Laboratory, operated under Contract No. DE-AC02-05CH11231 using NERSC award HEP-ERCAP0023978.
}

\bibliography{refs}

\let\addcontentsline\oldaddcontentsline% Restore \addcontentsline

\clearpage

\onecolumngrid
\begin{center}
  \textbf{\large Supplementary Material for Supernova axions convert to gamma-rays in magnetic fields of progenitor stars
  }\\[.2cm]
  \vspace{0.05in}
  {Claudio Andrea Manzari, Yujin Park, Benjamin R. Safdi, and Inbar Savoray}
\end{center}

\twocolumngrid

%%%%%%%%%% Merge with supplemental materials %%%%%%%%%%
\setcounter{equation}{0}
\setcounter{figure}{0}
\setcounter{table}{0}
\setcounter{section}{0}
\setcounter{page}{1}
\makeatletter
\renewcommand{\theequation}{S\arabic{equation}}
\renewcommand{\thefigure}{S\arabic{figure}}
\renewcommand{\thetable}{S\arabic{table}}

\setcounter{secnumdepth}{2}
\renewcommand{\thesection}{\Roman{section}}
\renewcommand{\thesubsection}{\thesection.\alph{subsection}}

\onecolumngrid

\startcontents[sections]
\tableofcontents

% \printcontents[sections]{l}{1}{\setcounter{tocdepth}{2}}

\section{Axion Emission from Supernovae}

\begin{figure*}[htb!]
	\centering
	\begin{tikzpicture}[very thick,k/.style={<-,semithick,yshift=5pt,shorten >=5pt,shorten <=5pt},snake/.style={decorate, decoration=snake},
		point/.style={
			thick,
			draw=black,
			cross out,
			inner sep=0pt,
			minimum width=6pt,
			minimum height=6pt,
		}, cross/.style={path picture={ 
			\draw[black]
			(path picture bounding box.south east) -- (path picture bounding box.north west) (path picture bounding box.south west) -- (path picture bounding box.north east);
	}}]
	
		\draw[snake] (0,0) -- (1.5,0) node[midway,below] {$\gamma$};
		\draw[dashed] (1.5,0) -- (3,0) node[midway,below] {$a$};
		\draw[snake] (1.5,0) -- (1.5,-1.5) node[pos=1.2, below] {$Z\,e$};
		
		\node [draw,circle,cross,minimum width=0.2 cm] at (1.5,-1.5){}; 
	\end{tikzpicture}
 \qquad 
 \begin{tikzpicture}[very thick,k/.style={<-,semithick,yshift=5pt,shorten >=5pt,shorten <=5pt}]
		\draw (0,0) -- (1.5,0) node[midway,below] {$n$};
        \draw (1.5,0) -- (3,0) node[midway,below] {$n$};
        \draw[dashed] (2.0,0) -- (3.0,0.75) node[shift={(-0.7,-0.2)}] {$a$};
		\draw[dashed] (1.5,0) -- (1.5,-1.5) node[midway, right] {$\pi$};
		
        \draw (0,-1.5) -- (1.5,-1.5) node[midway,below] {$n$};
        \draw (1.5,-1.5) -- (3,-1.5) node[midway,below] {$n$};
	\end{tikzpicture}
 \qquad \qquad 
 \begin{tikzpicture}[very thick,k/.style={<-,semithick,yshift=5pt,shorten >=5pt,shorten <=5pt}]
        \node[] (A) {};
        (A)
		\draw (0,0) -- (1.5,0) node[midway,below] {$n$};
        \draw (1.5,0) -- (3,0) node[midway,below] {$n$};
        \draw[dashed] (0,1.0) -- (1.5,0.) node[shift={(-1.,0.8)}, right] {$\pi$};
		\draw[dashed] (1.5,0) -- (3,1.0) node[shift={(-0.5,-0.2)}, left] {$a$};
\end{tikzpicture}\qquad\qquad
\begin{tikzpicture}[very thick,k/.style={<-,semithick,yshift=5pt,shorten >=5pt,shorten <=5pt}]
        \node[] (A) {};
        (A)
		\draw (0,0) -- (1.5,0) node[midway,below] {$n$};
        \draw (3.0,0) -- (4.5,0) node[midway,below] {$n$};
        \draw (1.5,0) -- (3.0,0) node[midway,below] {$n/\Delta$};
        \draw[dashed] (0,1.0) -- (1.5,0.) node[shift={(-1.,0.8)}, right] {$\pi$};
		\draw[dashed] (3.0,0) -- (4.5,1.0) node[shift={(-0.5,-0.2)}, left] {$a$};
\end{tikzpicture}
\caption{Diagrams contributing to axion production within a hot PNS from (left to right) Primakoff production, nucleon bremsstrahlung, pion-axion conversion, and pion-axion conversion with virtual nucleon or $\Delta$ exchange.  Diagrams related to these by symmetry are suppressed. }
\label{fig:diagrams}
\end{figure*}
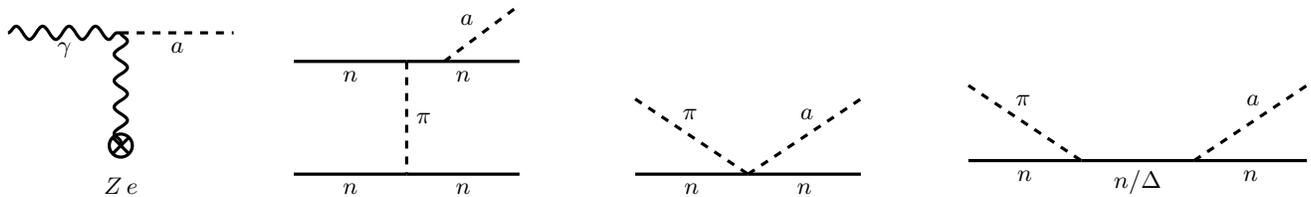

Axions are produced in hot PNSs through the diagrams illustrated in Fig.~\ref{fig:diagrams}. Here, we provide formulae for their emissivities, summarizing the results of Refs.~\cite{Raffelt:1985nk, Carenza:2019pxu, Ho:2022oaw}. Note that we account for the effects of degenerate media and  gravitational redshift~\cite{Buschmann:2019pfp,Caputo:2021rux}.\footnote{We do not account for the redshift from the radial velocity~\cite{Caputo:2021rux} because we estimate it has a subdominant effect on the axion luminosity and energy spectrum.}  

\subsection{Primakoff}

\begin{figure}[h!]
	\centering
	\begin{tikzpicture}[very thick,k/.style={<-,semithick,yshift=5pt,shorten >=5pt,shorten <=5pt},snake/.style={decorate, decoration=snake},
		point/.style={
			thick,
			draw=black,
			cross out,
			inner sep=0pt,
			minimum width=6pt,
			minimum height=6pt,
		}, cross/.style={path picture={ 
			\draw[black]
			(path picture bounding box.south east) -- (path picture bounding box.north west) (path picture bounding box.south west) -- (path picture bounding box.north east);
	}}]
	
		\draw[snake] (0,0) -- (1.5,0) node[midway,below] {$\gamma$};
		\draw[dashed] (1.5,0) -- (3,0) node[midway,below] {$a$};
		\draw[snake] (1.5,0) -- (1.5,-1.5) node[pos=1.2, below] {$Z\,e$};
		
		\node [draw,circle,cross,minimum width=0.2 cm] at (1.5,-1.5){}; 
	\end{tikzpicture}
\caption{Primakoff conversion of a photon into an axion in the electric field of a particle with charge $Ze$. }
\label{fig:Primakoff}
\end{figure}
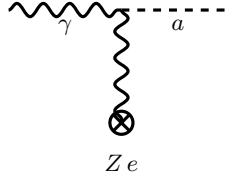

The two-photon coupling of the axion,
\begin{equation}
	\mathcal{L} \supset \frac{1}{4}g_{a\gamma\gamma} F_{\mu\nu} \tilde{F}^{\mu\nu} a = g_{a\gamma\gamma}\, \vec{E}\cdot \vec{B}\; a\,,
\end{equation}
allows for axion-to-photon conversion in the electric field of a spectator charged particle, as shown in Fig.~\ref{fig:Primakoff}.
The axion production rate per unit energy is 
\begin{equation}
	\frac{d\dot{N}_a}{d\omega} = 4\pi\int dR R^2\frac{d\dot{n}_a}{d\omega}\,,
\end{equation}
where the integral is over the radial direction in the PNS.  We define $\Gamma_{\gamma}(\omega)$ to be the Primakoff conversion rate, so that
\begin{equation}
	\frac{d\dot{n}_a}{d\omega} = \frac{\omega^2(1-\omega_p^2/\omega^2)}{\pi^2}\frac{1}{e^{\omega/T}-1}\Gamma_{\gamma}(\omega)\,,
\end{equation}
with $\omega_p$ the photon plasma frequency.
Ignoring recoil effects one finds 
\begin{align}
	\Gamma_{\gamma}(\omega) = \frac{g_{a\gamma\gamma}^2Z^2\, \alpha\, n^{\rm eff}}{8\pi}\int
d\Omega \frac{|\vec{k}_{\gamma}\times \vec{k}_a|^2}{|\vec{q}|^2(|\vec{q}|^2+\kappa^2)}\,,
\label{eq:GammaPrim}
\end{align}
where $n^{\rm eff}$ is the effective number density of the target, $Z$ is the atomic number of the target, $\vec{k}_{\gamma}$ and $\vec{k}_a$ are the momenta of the incoming photon and outgoing axion, and $\vec{q} = \vec{k}_{\gamma} - \vec{k}_a$ is the momentum transfer. Following Ref.~\cite{Raffelt:1985nk} we accounted for the plasma screening effect through the Debye factor $\kappa$. Expanding the momenta in~\eqref{eq:GammaPrim} up to $\mathcal{O}(\omega_p^2/\omega^2, m_a^2/\omega^2)$, we find
\begin{equation}
	\Gamma_\gamma(\omega) = \frac{g_{a\gamma\gamma}^2Z^2\, \alpha n^{\rm eff}(1-\xi)}{4\omega^2(2-\xi)^3}\bigg[-2\omega^2(2-\xi) + (\kappa^2+2(2-\xi)\omega^2)\log\bigg(\frac{\kappa^2+2\omega^2(2-\xi)}{\kappa^2}\bigg) \bigg]\,,
\end{equation}
where $\xi = \frac{\omega_p^2 + m_a^2}{\omega^2}$.\\

The Primakoff process is most relevant when the spectator particle is non-relativistic. In a PNS, the most important targets are thus protons. The protons are partially degenerate and their effective number density is given by~\cite{Payez:2014xsa}
\begin{equation}
	n_p^{\rm eff} = 2 \int \frac{d^3p}{(2\pi)^3}f_p(1-f_p)\,,
\end{equation}
where $f_p$ is the Fermi-Dirac distribution function for an interacting proton. Here we also note that there are finite temperature and density effects in a PNS that reduce the proton mass, enhancing its degeneracy. These effects are taken into account in our results. The Debye scale for the proton Coulomb potential is controlled by the longitudinal component of the polarization tensor~\cite{Raffelt:1985nk}. Using the one-loop, non-relativistic approximation for the polarization tensor gives~\cite{Payez:2014xsa} 
\begin{equation}
	\kappa^2 \simeq \frac{4\alpha m_p^*}{\pi}\int dp\, f_p = \frac{4\pi\alpha n_p^{\rm eff}}{T}\,,
\end{equation}
where $m_p^*$ is the effective proton mass. Finally, the plasma frequency in a degenerate medium as in a PNS is given by
\begin{equation}
	\omega_p^2 = \frac{4\pi\alpha n_e}{E_F} = \frac{4\alpha}{3\pi}p_F^2v_F \,,
\end{equation}
where $n_e$ is the electron number density and $E_F,\,p_F,\,v_F$ are the Fermi energy, momentum and velocity, respectively.  Note that $v_F \approx 1$ for the electron.

\subsection{Nucleon-Nucleon Bremsstrahlung}

\label{sec:nucelon_axion}

A relevant process of axion production in a PNS is the nucleon-nucleon (NN) axion bremsstrahlung, which involves the axion-nucleon coupling. In this work we follow Ref.~\cite{Carenza:2019pxu}, modelling this process including the effect of a massive pion propagator, the contribution of the $\rho$-meson exchange to mimic the effects of two-pion exchange, the medium modification of the nucleon mass, and accounting for nucleon multiple scatterings. 

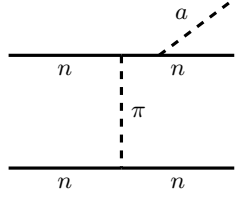
\begin{figure}[h!]
	\centering
	\begin{tikzpicture}[very thick,k/.style={<-,semithick,yshift=5pt,shorten >=5pt,shorten <=5pt}]
	
		\draw (0,0) -- (1.5,0) node[midway,below] {$n$};
        \draw (1.5,0) -- (3,0) node[midway,below] {$n$};
        \draw[dashed] (2.0,0) -- (3.0,0.75) node[shift={(-0.7,-0.2)}] {$a$};
		\draw[dashed] (1.5,0) -- (1.5,-1.5) node[midway, right] {$\pi$};
		
        \draw (0,-1.5) -- (1.5,-1.5) node[midway,below] {$n$};
        \draw (1.5,-1.5) -- (3,-1.5) node[midway,below] {$n$};
	\end{tikzpicture}
\caption{Nucleon-Nucleon Bremsstrahlung production of an axion. }
\label{fig:Bremsstrahlung}
\end{figure}

With the axion-nucleon interaction defined as
\begin{equation}
{\mathcal L} \supset \frac{C_{aNN}}{2f_a}\partial_{\mu}a\bar{N}\gamma^{\mu}\gamma^5 N\,,
\end{equation}
the axion production rate per unit energy is
\begin{align}
\frac{d\dot{n}_a}{d\omega} = \frac{\omega^3}{16 \pi^2}\frac{\Gamma_\sigma}{\omega^2+\Gamma^2}\frac{n_B}{f_a^2}\,  \bigg[s^{nn}(\omega/T)+s^{pp}(\omega/T)+s^{np}(\omega/T)\bigg] \,,
\end{align}
where $\Gamma_{\sigma}$ is the nucleon “spin fluctuation rate” and $s^{nn}(\omega/T)$, $s^{pp}(\omega/T)$ and $s^{np}(\omega/T)$ are 5-dimensional integrals, depending of the coefficients $C_{app}$ and $C_{ann}$, that can be found in Ref.~\cite{Carenza:2019pxu}, and $\Gamma$ accounts for many-body effects caused by multiple nucleon scatterings (see discussion in Ref.~\cite{Carenza:2019pxu}).

\subsection{Pion-Axion Conversion}
\label{sec:pion_axion}

\begin{figure}[h!]
\centering

\begin{tikzpicture}[baseline=(A),very thick,k/.style={<-,semithick,yshift=5pt,shorten >=5pt,shorten <=5pt}]
        \node[] (A) {};
        (A)
		\draw (0,0) -- (1.5,0) node[midway,below] {$n$};
        \draw (1.5,0) -- (3,0) node[midway,below] {$n$};
        \draw[dashed] (0,1.0) -- (1.5,0.) node[shift={(-1.,0.8)}, right] {$\pi$};
		\draw[dashed] (1.5,0) -- (3,1.0) node[shift={(-0.5,-0.2)}, left] {$a$};
\end{tikzpicture}\qquad\qquad
\begin{tikzpicture}[baseline=(A),very thick,k/.style={<-,semithick,yshift=5pt,shorten >=5pt,shorten <=5pt}]
        \node[] (A) {};
        (A)
		\draw (0,0) -- (1.5,0) node[midway,below] {$n$};
        \draw (3.0,0) -- (4.5,0) node[midway,below] {$n$};
        \draw (1.5,0) -- (3.0,0) node[midway,below] {$n/\Delta$};
        \draw[dashed] (0,1.0) -- (1.5,0.) node[shift={(-1.,0.8)}, right] {$\pi$};
		\draw[dashed] (3.0,0) -- (4.5,1.0) node[shift={(-0.5,-0.2)}, left] {$a$};
\end{tikzpicture}

\caption{Axion-pion conversion processes in a SN through a contact interaction or the exchange of a virtual nucleon or $\Delta$ baryon.}
\label{fig:NucleonPion}
\end{figure}
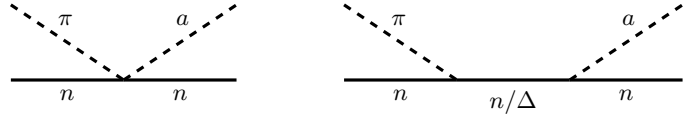

The Feynman diagrams for pion conversion are shown in Fig.~\ref{fig:NucleonPion}.
The axion emission rate per unit energy is~\cite{Ho:2022oaw,Carenza:2023lci}
\begin{equation}
    \begin{split}
    \frac{d\dot{n}_a}{d\omega} =& \mathcal{C}_a^{p\pi^-}\, \frac{(m_N T)^{3/2}}{2^{3/2}\pi^5 f_a^2}\left(\frac{m_N}{2f_\pi}\right)^2\left(\omega^2-m_a^2\right)^\frac{3}{2} \,\frac{\Theta(\omega-\max\left(m_{a},m_{\pi}\right))}{\exp{\left(x_a-y_\pi-\hat{\mu}_\pi\right)}-1}\,\frac{(\omega^2-m_\pi^2)^\frac{1}{2}}{\omega^{2}+\Gamma^2}\\
    &\times\int_0^\infty du\, u^2\frac{1}{\exp{\left(u^2-\hat{\mu}_p\right)}+1}\frac{1}{\exp{\left(-u^2+\hat{\mu}_n\right)}+1}\,,
    \end{split}
    \label{eq:QaPionDef}
\end{equation}
where $y_{\pi}=m_{\pi}/T$, $\hat{\mu}$ is the degeneracy parameter for pions and nucleons, $\Gamma$ accounts for many-body effects caused by multiple nucleon scatterings, and the Heaviside theta function fixes the minimal threshold energy.
The coefficient $\mathcal{C}_a^{p\pi^-}$, at first order in $1/m_N$, is given by~\cite{Ho:2022oaw}
\begin{align}
\begin{split}
\mathcal{C}_a^{p\pi^-} & =
\frac{2 g_{A}^{2}\big(2C_{+}^2 + C_{-}^2\big)}{3}
\bigg(\frac{\sqrt{\omega^2-m_{\pi}^2}}{m_N}\bigg)^{2}
+C_{a\pi N}^{2}\bigg(\frac{\omega}{m_{N}}\bigg)^{2}+\frac{g_{A}^{2}C_{aN\Delta}^2 }{9}\mathcal{F}_{a\pi N}
\bigg(\frac{\sqrt{\omega^2-m_{\pi}^2}}{m_N}\bigg)^{2}\\
&\quad-\frac{4\sqrt{3} g_A^{2} C_{aN\Delta} }{9}
\mathcal{F}_{a N\Delta}
\bigg(\frac{\sqrt{\omega^2-m_{\pi}^2}}{m_N}\bigg)^{2}\,,
\end{split}
\end{align}
with
\begin{align}
\begin{split}
\mathcal{D}&=\big[ 
\big(\Delta m - E_{\pi}\big)^{2} + \Gamma_\Delta^2 /4\big]\big[ 
\big(\Delta m + E_{\pi}\big)^{2} + \Gamma_\Delta^2 /4\big]\,,\quad \mathcal{F}_{a\pi N}=\mathcal{D}^{-1}E_{\pi}^2 \Big(\Delta m^2 + 2 E_{\pi}^2 + \frac{\Gamma_\Delta^2}{4} \Big)
\,,\\
\mathcal{F}_{a N\Delta}&=\mathcal{D}^{-1}E_{\pi}
\Big[\big(\Delta m^2 - E_{\pi}^2 \big) \big(C_{+} \Delta m + C_{-} E_{\pi}\big) 
+ \frac{\Gamma_\Delta^2}{4} \big(C_{+} \Delta m - C_{-}E_{\pi} \big) \Big]\,,
\end{split}
\end{align}
where $C_+ = C_{app}+C_{ann}$, $C_- = C_{ann}-C_{app}$, $\Gamma_{\Delta}=117$ MeV is the width of the $\Delta$-resonance, $\Delta m=m_{\Delta}-m_{N}^{*}$ and $g_A\simeq 1.28$ is the axial-vector coupling constant. The axion-nucleon-pion and axion-nucleon-$\Delta$ couplings are defined in the next section.

Note that referring to {\it e.g.} Fig.~\ref{fig:luminosities} the pion-conversion processes give non-trivial emission up to high energies $\sim$1 GeV.  On the other hand, the heavy baryon chiral EFT is not expected to be valid for axion energies of order the nucleon mass (around 1 GeV). With that said, we stress that due to the steeply falling spectral shape with energy, the axions with energies around 1 GeV make a negligable contribution to our final result. For example, only including axion energies less than 500 MeV changes our projected upper limits by less than 1\%.  Additionally, we note that in some regions of the PNS our simulations predict a pion condensate ($\mu_{\pi^-} > m_{\pi^-}$). In this work we do not attempt to correctly describe this phase or to account for nucleon-pion interactions that can affect the number density~\cite{Carenza:2020cis}. 
 On the other hand, we verify that removing the small regions that give pion condensates leads to an imperceptible
 change in our final results.  Still, it is the pion condensate that gives rise to the sharp spectral features seen in {\it e.g.} Fig.~\ref{fig:TimeDistr}. 

\section{Axion-matter couplings under RG flow}

In this section we consider axion-like particle scenarios where in the UV the axion only couples to electroweak gauge bosons.  As is well established~\cite{Srednicki:1985xd,Bauer:2017ris}, the axion-matter couplings are generated under the RG flow.  We review the expected loop-induced couplings here.

By assumption, in the UV the axion only couples to the Standard Model through the terms
\begin{equation}
    \mathcal{L} \supset -\frac{1}{4}C_{aWW} \frac{g_2^2}{8\pi^2 f_a} a\, W_{\mu\nu}^a\tilde{W}^{a\, \mu\nu} - \frac{1}{4}C_{aBB} \frac{g_1^2}{8\pi^2 f_a} a\, B_{\mu\nu}\tilde{B}^{\mu\nu} \,,
\end{equation}
with $W_{\mu \nu}$ and $B_{\mu \nu}$ the $SU(2)_L$ and $U(1)_Y$ field strengths, respectively, and
where by assumption we do not allow the axion-like particle to couple to QCD.
Note that the coupling constants $g_1$ and $g_2$ are not topological quantities and therefore are not scale invariant, while $C_{aWW}$ and $C_{aBB}$ do not evolve under RG flow. At the electroweak symmetry breaking scale, we integrate out the heavy fields, and we are left with the effective Lagrangian 
\begin{equation}
    \mathcal{L} \supset - \frac{1}{4}C_{a\gamma\gamma} \frac{e^2}{8\pi^2 f_a} a\, F_{\mu\nu}\tilde{F}^{\mu\nu}\,,
\end{equation}
where $F_{\mu\nu}$ is the electromagnetic field tensor and the matching simply reads $C_{a\gamma\gamma} = C_{aWW} + C_{aBB}$.

\subsection{Couplings with Nucleons}

On the other hand, the couplings of an axion to fermions are not topological and are induced through RG running. Recall that the coefficient $C_{aqq}$ describing the axion coupling to quark fields $q$ is defined by 
\begin{equation}
    \mathcal{L} \supset \frac{C_{aqq}}{2f_a}\partial^{\mu} a\, \bar{q}\gamma_{\mu}\gamma_5\, q\,.
    \label{eq:quarkLag}
\end{equation}
Then,
at one-loop we have (see, {\it e.g.},~\cite{Bauer:2020jbp})
\begin{equation}
    \frac{d C_{aqq}(\mu)}{d\ln{\mu}} = -\frac{3}{128\pi^4}\left(\frac{3}{4}g_2^4C_{aWW} + (\mathcal{Y}_Q^2+\mathcal{Y}_q^2)g_1^4C_{aBB}\right)\,.
\end{equation}
where $\mathcal{Y}_Q$ and $\mathcal{Y}_q$ are the hypercharges of the left-handed quark doublet and right-handed singlet, respectively.\\ 
Including the one-loop RG flow for the gauge coupling
\begin{equation}
    \frac{dg_1}{d\ln\mu} = \frac{41}{6}\frac{g_1^3}{16\pi^2}\,, \qquad \frac{dg_2}{d\ln\mu} = -\frac{19}{6}\frac{g_2^3}{16\pi^2}\,,
\end{equation}
we obtain
\begin{equation}
    C_{aqq}(\mu) = \frac{3}{128\pi^4}\ln\left(\frac{\Lambda}{\mu}\right)\bigg(\frac{3}{4}C_{aWW}g_2(\Lambda)^2g_2(\mu)^2 + (\mathcal{Y}_Q^2+\mathcal{Y}_q^2)C_{aBB}g_1(\Lambda)^2g_1(\mu)^2 \bigg)\,,
    \label{eq:RunningtoEW}
\end{equation}
with $\Lambda \sim 4 \pi f_a$ the UV scale where, by construction, $C_q(\Lambda) = 0$.
From~\eqref{eq:RunningtoEW} we can compute $C_q(\mu = m_z)$ at the EW scale, given by the $Z$-boson mass $m_z$. Below the electroweak scale the RG evolution proceeds only through the contribution of $U(1)_{\rm EM}$:
\begin{equation}
    \frac{d C_{aqq}(\mu)}{d\ln{\mu}} = -\frac{3}{64\pi^4}Q^2e^4C_{a\gamma\gamma}\,,
\end{equation}
with $Q$ the electromagnetic charge of the quark.
(Note that we neglect the running of the fine structure constant as we expect this contribution to be subdominant.) 
In the IR, at $\Lambda_{\rm QCD} < \mu < m_z$, we thus find
\begin{equation}
    C_{aqq}(\mu) = C_{aqq}(m_Z) + \frac{3}{64\pi^4}\ln\left(\frac{m_Z}{\mu}\right) Q^2e^2(m_Z)e^2(\Lambda_{\rm QCD})C_{a\gamma\gamma}\,.
    \label{eq:RunningBelowEW}
\end{equation}

Following Ref.~\cite{GrillidiCortona:2015jxo}, we extract the couplings of an axion to nucleons matching onto an effective Lagrangian valid at energies much lower than the QCD mass gaps $\sim\mathcal{O}(100\, \rm{MeV})$. The matching scale is conveniently chosen to be $\sim 2$ GeV, and we obtain
\begin{align}
{\cal L}_N  =\bar N v^\mu D_\mu N
+\frac{\partial_\mu a}{f_a} \Bigl \{ &
\frac{C_{auu}-C_{add}}2 (\Delta u-\Delta d)\bar N \frac{\gamma^\mu\gamma^5}{2} \sigma^3  N  \nonumber \\
&  +\Bigl[ \frac{C_{auu}+C_{add}}{2}(\Delta u+\Delta d) 
	+ \sum_{q=s,c,b,t} C_{aqq} \Delta q \Bigr] \bar N \frac{\gamma^\mu\gamma^5}{2} N
\Bigr \} \,,
\end{align}
where $C_{aqq} = C_{aqq}(\mu = 2\, \rm{GeV})$ and $N=(p,\,n)$. From low energy experiments and lattice determinations the matrix elements read~\cite{GrillidiCortona:2015jxo}
\begin{equation}
    \Delta u(2\, {\rm{GeV}}) = 0.897(27),\quad \Delta d(2\, {\rm{GeV}}) = -0.376(27),\quad \Delta s(2\, {\rm{GeV}}) = -0.026(4)\,,
\end{equation}
and the contribution from heavier quarks can be safely neglected. In analogy with~\eqref{eq:quarkLag}, the interaction terms can be rewritten as
\begin{align}
{\mathcal L} \supset \frac{\partial_\mu a}{2f_a} \Bigl[
 C_{app} \bar p \gamma^\mu\gamma^5 p  +C_{ann} \bar n \gamma^\mu\gamma^5 n
\Bigr] \,,
\label{eq:Lagnucleons}
\end{align}
with
\begin{equation}
\begin{split}
    &C_{app} = 0.88(3)C_u -0.39(2)C_d -0.038(5)C_s\,,\\
    &C_{ann} = 0.88(3)C_d -0.39(2)C_u -0.038(5)C_s\,.
\end{split}
\label{eq:NuclCoupl}
\end{equation}
Note that we neglect the self-renormalization of the fermionic axial current, which is sub-leading to the effect induced by the photon coupling, since we assume $C_{aqq}(\Lambda) = 0$ in the UV.

\subsection{Axion-Nucleon-Pion and Axion-Nucleon-$\Delta$ Couplings}

The axion-pion-nucleon coupling can be described in heavy baryon chiral perturbation theory~\cite{Chang:1993gm,Vonk:2021sit} and reads 
\begin{align}
    \mathcal{L}_{a\pi N} = i \frac{\partial_{\mu} a}{2 f_a} C_{a\pi N} (\pi^+ \bar{p}\gamma^{\mu} n - \pi^{-}\bar{n}\gamma^{\mu}p)\,,
\end{align}
with $C_{a\pi N} = (C_{app}-C_{ann})/{\sqrt{2}g_A}$.
Similarly, one can compute the axion-nucleon-$\Delta$ interaction~\cite{Ho:2022oaw} as 
\begin{align}
    \mathcal{L}_{aN\Delta} = i \frac{\partial_{\mu a}}{2 f_a} C_{aN\Delta}\bigg[ \bar{p}\,\Delta^{\mu\, +} + \bar{\Delta}^{\mu\, +}\,p + \bar{n}\,\Delta^{\mu\, 0} + \bar{\Delta}^{\mu\, 0}\,n \bigg]\,,
\end{align}
where $C_{aN\Delta} = -\sqrt{3}(C_{app}-C_{ann})/2$.

\subsection{Numerical Results}

Taking the UV scale to be $\Lambda = 10^9\, {\rm GeV}$,  then~\eqref{eq:RunningBelowEW} together with~\eqref{eq:NuclCoupl} gives
\begin{equation}
\begin{split}
    &C_{app} \sim 3.5\times 10^{-5}\, C_{aBB} + 2.1\times 10^{-4}\, C_{aWW}\,,\\
    &C_{ann} \sim -5.6 \times 10^{-6}\, C_{aBB} + 2.1\times 10^{-4}\,C_{aWW}\,.
\end{split}
\end{equation}
For a generic UV completion where $C_{aWW} \sim C_{aBB}$, we thus expect $C_{app} / C_{a\gamma\gamma} \sim C_{ann} / C_{a\gamma\gamma} \sim 10^{-4}$, $C_{a\pi N} / C_{a\gamma\gamma} \sim 10^{-5}$, and $C_{a N \Delta} / C_{a\gamma\gamma} \sim 2 \times 10^{-5}$, justifying the benchmark values assumed for an axion-like particle model in the main Letter.

\section{KSVZ and DFSZ Benchmarks}

Throughout this Letter, we show our exclusion limits and projections for the KSVZ and DFSZ axion benchmarks. Here, we briefly summarize the coupling strengths expected for these models (see, {\it e.g.},~\cite{DiLuzio:2020wdo} for a more complete review). 

Recall that in the KSVZ axion model, the Standard Model is extended by a complex scalar and a vector-like fermion that interact through a Yukawa interaction. The scalar is a singlet under the SM gauge groups while the fermion is in the fundamental representation of $SU(3)_C$.  Both fields are charged under a global $U(1)_{\rm PQ}$ symmetry. In this minimal setup, the axion-photon coupling originates purely from axion-pion mixing in the IR, while the couplings of the axion to nucleons arise from the axion coupling to gluons. The couplings of the axion to the photon, neutron and proton are~\cite{GrillidiCortona:2015jxo}:
\begin{align}
    C_{a\gamma\gamma} \approx -1.92\,,\quad C_{ann} \approx -0.02 \,,\quad C_{a pp} \approx -0.47 \,.
\end{align}

Conversely, in the DFSZ setup, the Standard Model quark fields are charged under $U(1)_{PQ}$ and the scalar content of the SM has to be extended by one singlet and at least one additional Higgs doublet. The axion-photon coupling is therefore modified by the presence of an electromagnetic anomaly, and the couplings to quarks have a tree-level contribution. For the extensions with a single additional Higgs doublet, the couplings to the photon, neutron and proton are
\begin{align}
    C_{a\gamma\gamma} \approx 0.75 \,,\quad C_{a pp}\approx (-0.617+0.435\cos^2\beta)\,,\quad C_{a nn}\approx(0.254-0.414\cos^2\beta)\,, 
\end{align}
where $\tan\beta$ is the ratio between the vacuum expectation values of the two Higgs doublets in the theory. In this work, we present limits for the DFSZ scenario as a band obtained varying the value of $\tan{\beta}$ (see, {\it e.g.}, Fig.~\ref{fig:SN_limit_projs}).

\section{Magnetic field modeling and axion-photon conversion}

Here, we briefly summarize how we compute the axion-to-photon conversion probabilities.
Assuming the axion and photon wavelengths are much smaller than the propagation length, the second-order axion-photon mixing equations can be linearized to first-order mixing equations~\cite{Raffelt:1987im}. Hence, the equation of motion for a mode of energy $\omega$ that propagates in the z-direction through an external magnetic field $\vec{B}(\vec{x})$, in a plasma with electron density $n_e(\vec{x})$, reads
\begin{equation}
	\bigg( (\omega -i\partial_z)\mathbb{I} + M(z)\bigg)
	\begin{pmatrix}
		|\gamma_\perp(z)\rangle\\
		|\gamma_\parallel(z)\rangle\\
		|a(z)\rangle\\
	\end{pmatrix} = 0\,,
	\label{eq:EqofMotion}
\end{equation}
where $|\gamma_{\parallel(\perp)}\rangle$ represents an electromagnetic wave with the electric field oscillating in the direction parallel (perpendicular) to the projection of $\vec{B}(\vec{x})$ on the $x-y$ plane $\vec{B}_T$, and
\begin{equation}
	M(z) = 
	\begin{pmatrix}
		-\frac{\omega_p^2(\vec{x})}{2\omega} + \delta^\perp_{\rm HE}(\vec{x})&0 &0\\
		0 &-\frac{\omega_p^2(\vec{x})}{2\omega} + \delta^\parallel_{\rm HE}(\vec{x}) &g_{a\gamma\gamma}\frac{B_T(\vec{x})}{2}\\
		0 &g_{a\gamma\gamma}\frac{B_T(\vec{x})}{2} &-\frac{m_a^2}{2\omega}\\
	\end{pmatrix}\,.
\end{equation}
Here,  $\omega_p(\vec{x}) = \sqrt{4\pi e^2 n_e(\vec{x})/m_e}$ is the plasma frequency for a non-degenerate medium ({\it e.g.}, the electron interstellar plasma or the exterior plasma of a star), and the Euler-Heisenberg term, $\delta^{\parallel}_{\rm HE}(\vec{x}) = 7\alpha_{\rm em}\omega /(90\pi B_{\rm crit}^2) |\vec{B}(\vec{x}) \times \hat{z}|^2\,,\delta^{\perp}_{\rm HE}(\vec{x}) = 4/7\delta^{\parallel}_{\rm HE}(\vec{x})$\, with $B_{\rm crit} = 4.41 \times 10^{13}$ G, accounts for strong-field QED effects in vacuum~\cite{heisenberg2006consequences}. We neglect the Faraday rotation related to the longitudinal component of the external field.
Since the axion field only couples to $\gamma_\parallel$, axion-photon conversion can be described by the bottom $2\times2$ sub-block of $M$, which we denote as $M_{a\gamma_\parallel}$. The solution of equation (\ref{eq:EqofMotion}), after a distance R, is then given by the path-ordered transfer matrix
\begin{equation}
	\label{eq:homsoln}
	\begin{pmatrix}
		|\gamma_\parallel(R)\rangle\\
		|a(R)\rangle\\
	\end{pmatrix} =  {\cal P}_z\left[ \exp \left(-i \omega R\, \mathbb{I} \ -i \int_0^R M_{a\gamma_\parallel}(z) d z\right) \right]
	\begin{pmatrix}
		|\gamma_\parallel(0)\rangle\\
		|a(0)\rangle\\
	\end{pmatrix}\,.
\end{equation}
Throughout this work we compute this conversion probability numerically using the expression above.

\subsection{Galactic conversion probability}
\label{sec:gal}

As part of this work we reinterpret the SMM data in terms of axion-photon conversion in the Galactic magnetic fields. We use both the parametric Jansson \& Farrar (JF) model~\cite{Jansson:2012pc}, as used in previous works~\cite{Payez:2014xsa,Hoof:2022xbe}, along with the newer Unger \& Farrar (UF) models~\cite{Unger:2023lob}.  Additionally, we use the \texttt{ne2001} free-electron density model for the Galaxy~\cite{Cordes:2002wz}, though we find this is a minor correction; setting $n_e = 0$ throughout the Galaxy changes our Galactic limits by less than $0.1$\%. We note that UF present 8 variants of their Galactic field model to account for modeling uncertainties; in Fig.~\ref{fig:Bvariation} we shade the difference in $g_{a\gamma\gamma}$ upper limits found using this ensemble of magnetic field models. We also compare these results to that found using the JF model; the difference is relatively minor. In Fig.~\ref{fig:SN_limit_projs} we, at each mass $m_a$, chose the weakest upper limit from the ensemble of UF models.

\subsection{Red supergiant conversion probability}
\label{sec:RSG}

In the main Letter we focus on BSG supernovae since that was the case for SN1987A. BSGs are relatively compact supergiants with strong dipole magnetic fields.  In contrast, most core collapse supernovae are in fact expected to arise from RSG progenitors. There are important differences between RSGs and BSGs. In particular, typical RSGs are around an order of magnitude larger in radius than BSGs; accordingly, the dipole magnetic fields for RSGs are much smaller than those of BSGs.  Here, we briefly summarize our modeling of the conversion probability for RSGs.  Our predicted upper limit for a Galactic RSG SN is shown in Fig.~\ref{fig:SimVariation_RSG}, which is to be compared with the BSG projections in Fig.~\ref{fig:SN_limit_projs}; all aspects apart from the conversion probability are the same between these two projections. 

Little is known about the surface magnetic field distribution and geometry for RSGs.  Many RSGs have surface-averaged, longitudinal magnetic fields $\sim$1 G (see, {\it e.g.}, Betelgeuse~\cite{2010A&A...516L...2A}).  On the other hand, this does not mean that at any point on the RSG surface we expect $B \sim 1$ G. Rather, it is plausible, though not definitive, that the surface-averaged longitudinal magnetic field measurements are explained by large convection cells within the outer layers of the RSG, which generate magnetic fields through the dynamo effect.  For example, Ref.~\cite{Dorch:2004af} performed magneto-hydrodynamic simulations of RSGs with parameters similar to Betelgeuse and reproduced the observed surface-averaged longitudinal field strength of around 1 G but, at the same time, found that most individual locations on the stellar surface have $|B| > 50 $ G with some points having $|B| > 500$ G; the signed-average of the longitudinal field, which is what is measured, averages to the lower quantity. 

Without access to modern magneto-hydrodynamic simulations of RSGs, we chose to model the magnetic field distribution, roughly, using the equipartition theorem. (Note that the fields found in~\cite{Dorch:2004af} actually exceeded the equipartition-strength fields by around a factor of two.)  In particular, we apply an equipartition argument to estimate the field strengths generated by the dynamo effect in the large convection cells in the simulations performed in Ref.~\cite{2022ApJ...929..156G}. (Note that Ref.~\cite{2022ApJ...929..156G} does not itself keep track of the magnetic fields.)  That work simulates the hydrodynamics of the outer $\sim$30\%, by volume, of RSG-type stars.  They perform two simulations,  RSG1L4.5 with a photosphere radius $R_{\rm phot} \approx 800$ $R_\odot$, and RSG2L4.9, with $R_{\rm phot} \approx 900$ $R_\odot$.  Both simulations yield similar results for the axion-to-photon conversion probability, but we adopt RSG1L4.5 for definiteness (the upper limits at low $m_a$ found using RSG2L4.9 are less than 20\% stronger).  Note that both simulations find large convection cells, with correlation lengths larger than 100 $R_\odot$.  

\begin{figure}[h]
    \includegraphics[width=0.3\columnwidth]{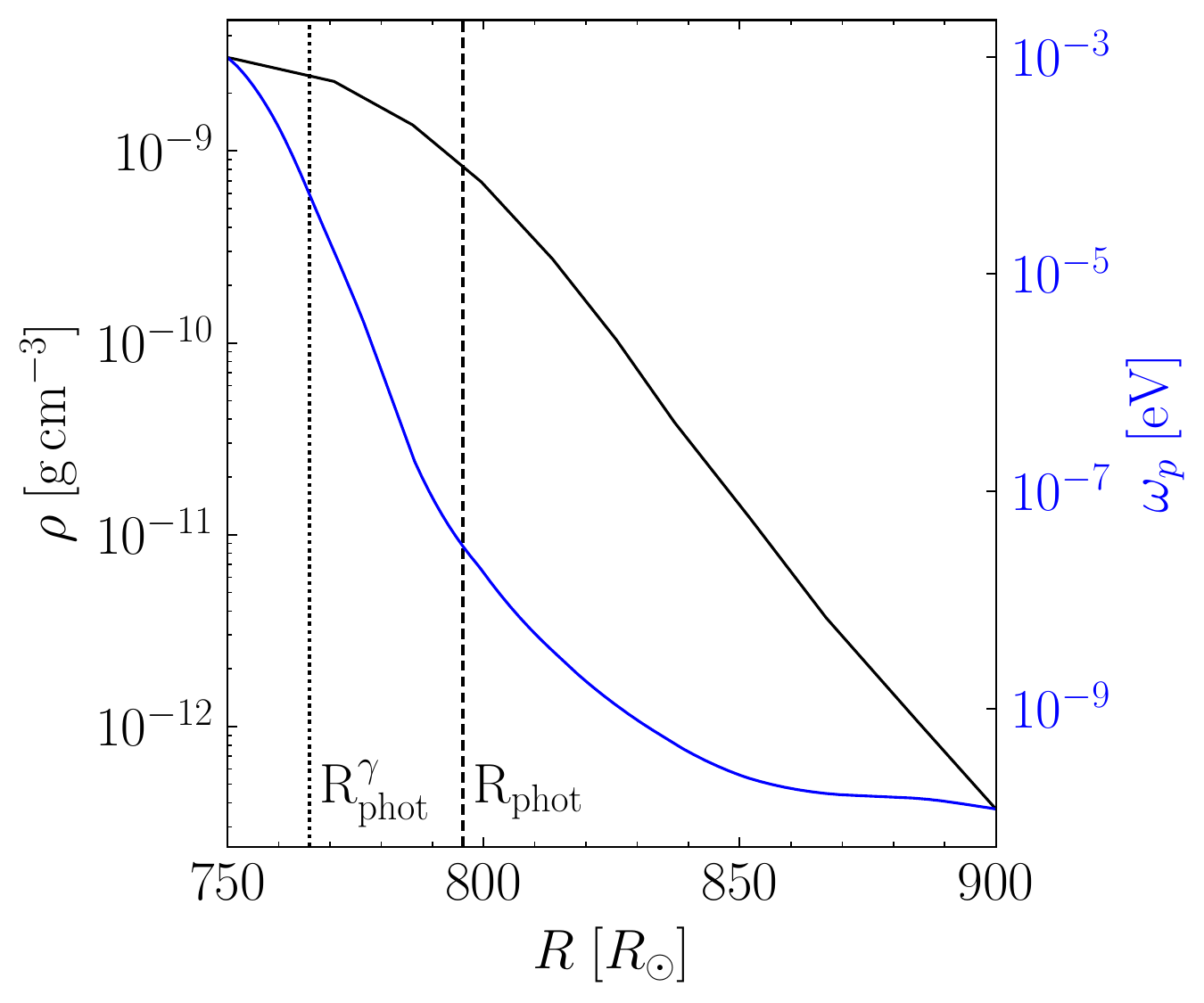}\includegraphics[width=0.3\columnwidth]{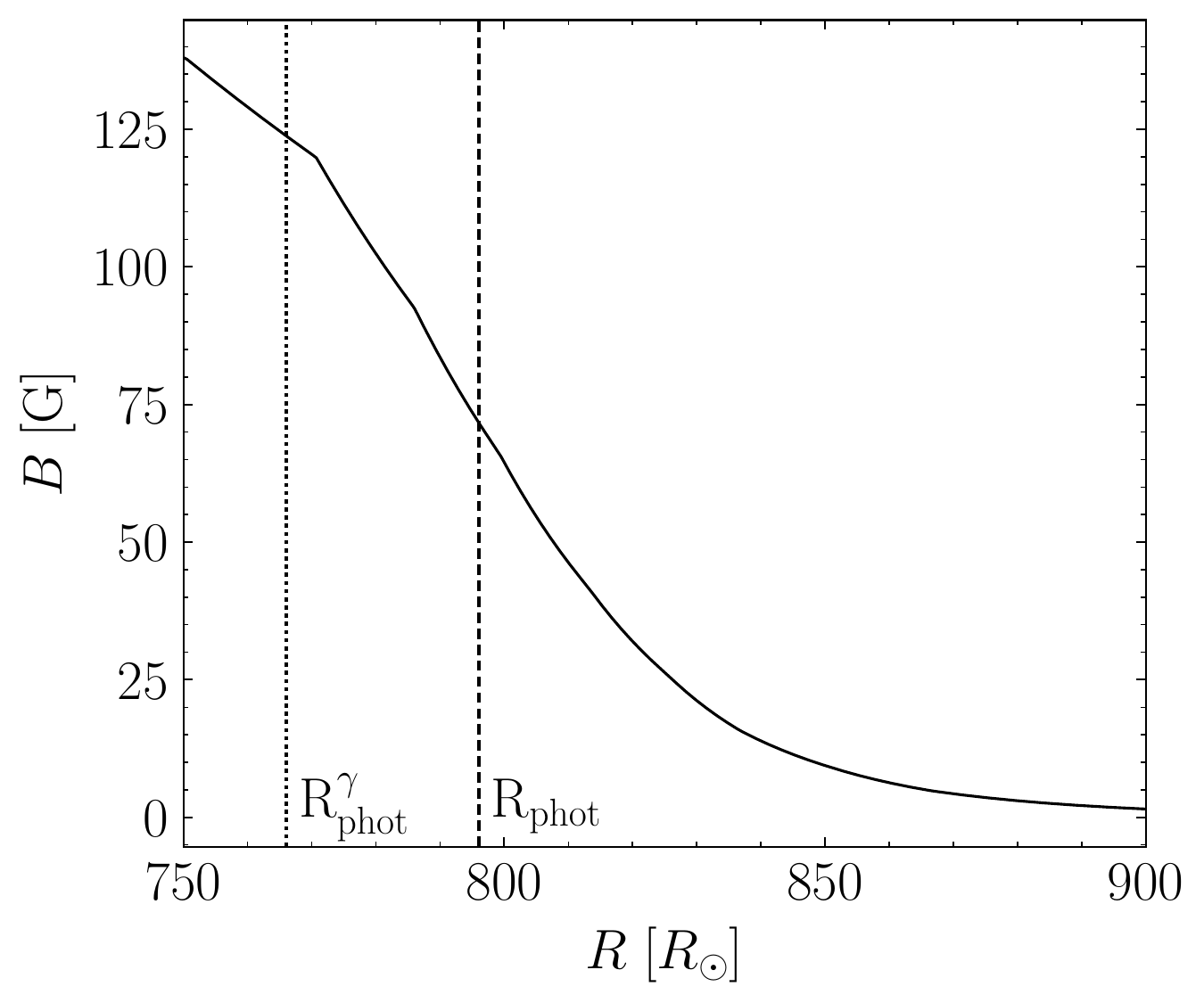}
    \includegraphics[width=0.3\columnwidth]{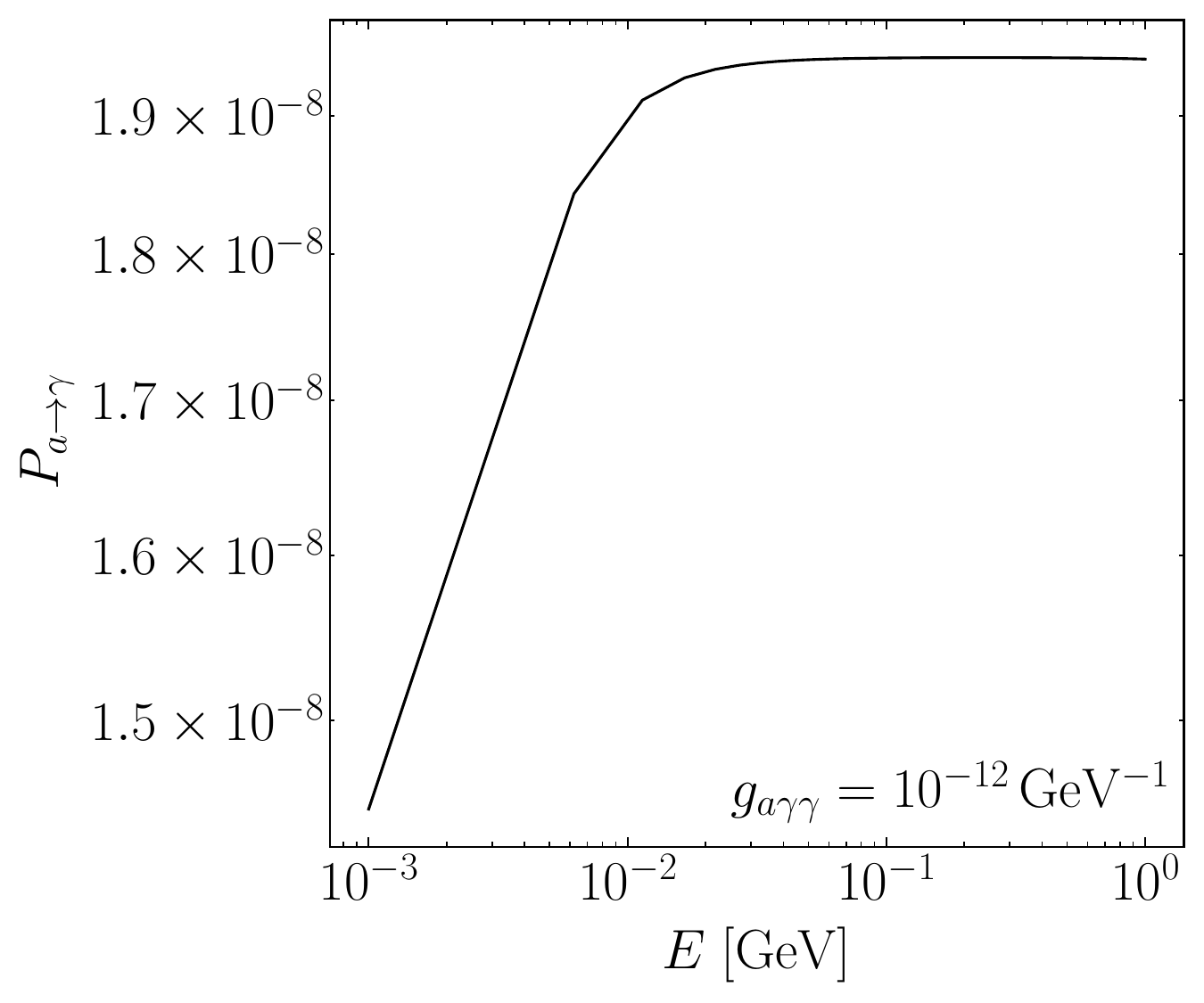}
    \caption{ (Left) Average density and plasma frequency from the RSG hydrodynamic simulation RSG1L4.5 in Ref.~\cite{2022ApJ...929..156G}. We mark the photosphere radius and the radius from which 100 MeV gamma-rays typically free-stream outwards without scattering, $R_{\rm phot}^\gamma$. (Center) Our inferred magnetic field strength as a function of the radius $R$ within the star, assuming equipartition between the magnetic energy density and the kinetic energy density in convection. (Right) The axion-to-photon conversion probability (for low axion masses $m_a \sim 0$ eV) as a function of the axion energy $E$, assuming $g_{a\gamma\gamma} = 10^{-12}$ GeV$^{-1}$ for illustration.
    }
    \label{fig:RSG_plots}
\end{figure}

We apply the equipartition theorem to equate the kinetic energy density in the plasma, $E_{\rm kin} = {1\over 2} \rho \langle v^2 \rangle$, to the magnetic field energy density $E_B = {1 \over 2} B^2$, giving $B = \sqrt{ \rho \langle v^2 \rangle }$, with $\rho$ the average density and $\langle v^2 \rangle$ the average velocity of the convection cells.  Ref.~\cite{2022ApJ...929..156G} finds $\sqrt{\langle v^2 \rangle} \sim 10 \, \, {\rm km/s}$ roughly independent of radius, though we adopt $\sqrt{\langle v^2 \rangle} = 7$ km/s for definiteness and to be conservative.  We extract the average density as a function of radius $R$ from Ref.~\cite{2022ApJ...929..156G}; we reproduce this curve in Fig.~\ref{fig:RSG_plots}, where we also indicate $R_{\rm phot}$ and the radius (interior to $R_{\rm phot}$) where the star becomes transparent to gamma-rays ({\it i.e.}, 2/3 of gamma-rays escape from this radius with no scattering for $E = 100$ MeV. The energy dependence is negligible within the range of interest for this work). We refer to this gamma-ray transparency radius as $R_{\rm phot}^\gamma$.  Note that Ref.~\cite{2022ApJ...929..156G} also provides the average temperature as a function of radius. Assuming thermal equilibrium we may then calculate the photon plasma frequency $\omega_p$ as a function of $R$, which is also shown in Fig.~\ref{fig:RSG_plots}.  Fig.~\ref{fig:RSG_plots} also shows our derived equipartition $B$, which is on the order of 100 G, consistent with~\cite{Dorch:2004af}.

We compute the axion-to-photon conversion probabilities by assuming radial trajectories with the mean quantities given above, beginning at $R_{\rm phot}^\gamma$. (Note, however, that given the large values of $\omega_p$, our results are numerically stable to starting the trajectories at arbitrary points further into the RSG interiors.) We project the expected values of $B$ along the transverse direction, but we assume a single domain, given that Ref.~\cite{2022ApJ...929..156G} finds large domains but does not present more quantitative information about the domain size that could be used to refine this estimate. In Fig.~\ref{fig:RSG_plots} we show the energy-dependent axion-to-photon conversion probabilities.

Using the conversion probabilities described above, we project the sensitivity to a next Galactic SN assuming a RSG progenitor, with the results shown in Fig.~\ref{fig:SimVariation_RSG} under identical conditions otherwise to that assumed for the BSG SN in Fig.~\ref{fig:SN_limit_projs}.  The RSG sensitivity is similar to but slightly worse than that for a BSG, though we caution that our RSG conversion probability calculations are much less robust than those we present for BSG, which we model simply as vacuum dipoles. It is interesting and important to further refine the axion-to-photon conversion probability calculation in RSGs by using dedicated magneto-hydrodynamic simulations; we leave this to future work.  It is also possible that the SN simulations themselves should be modified for RSGs relative to our fiducial choice. 

\section{SN Simulations: Intermediate Quantities}\label{sec:SNSim}

\begin{figure}[h]
    \includegraphics[width=0.3\columnwidth]{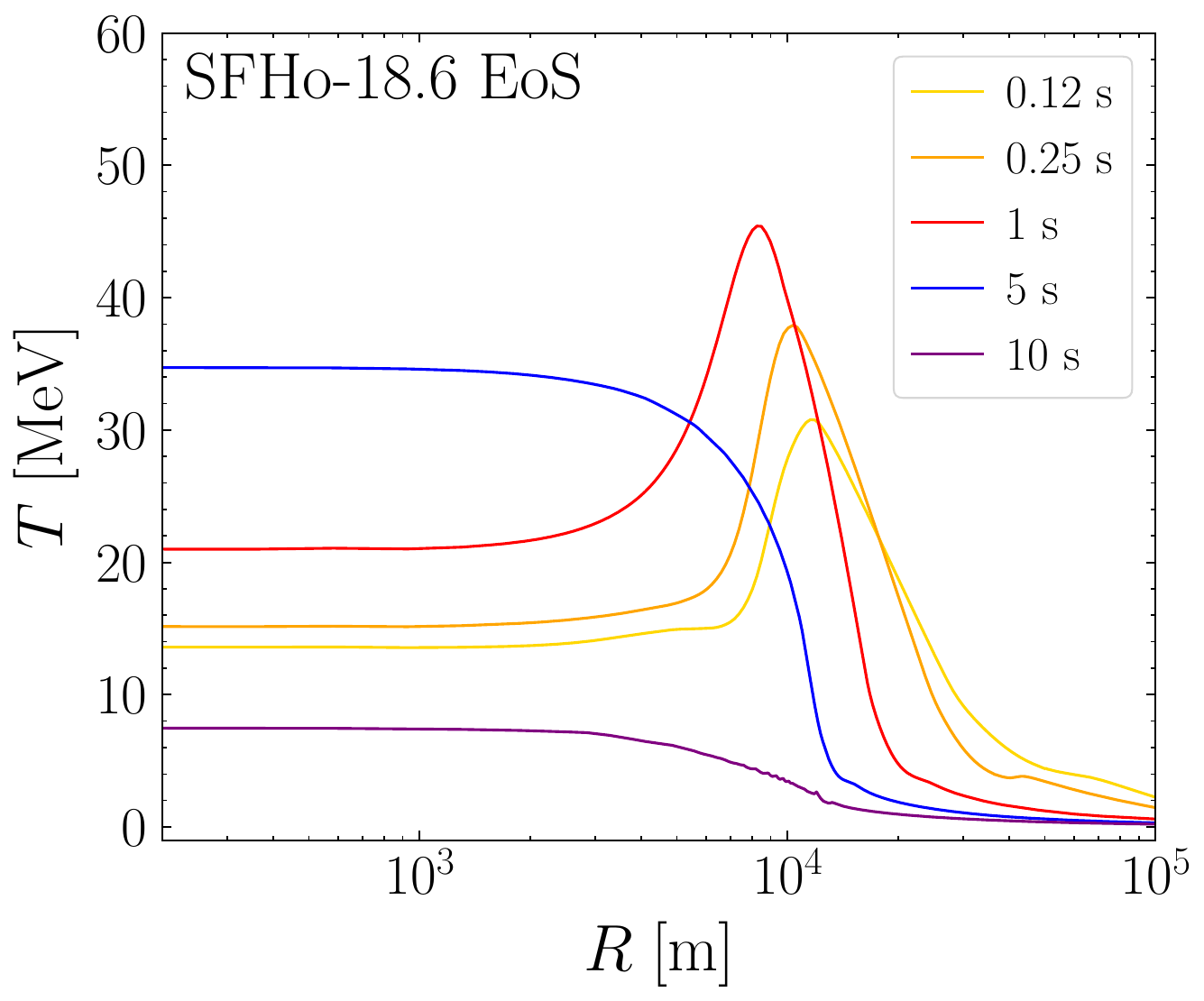} \includegraphics[width=0.3\columnwidth]{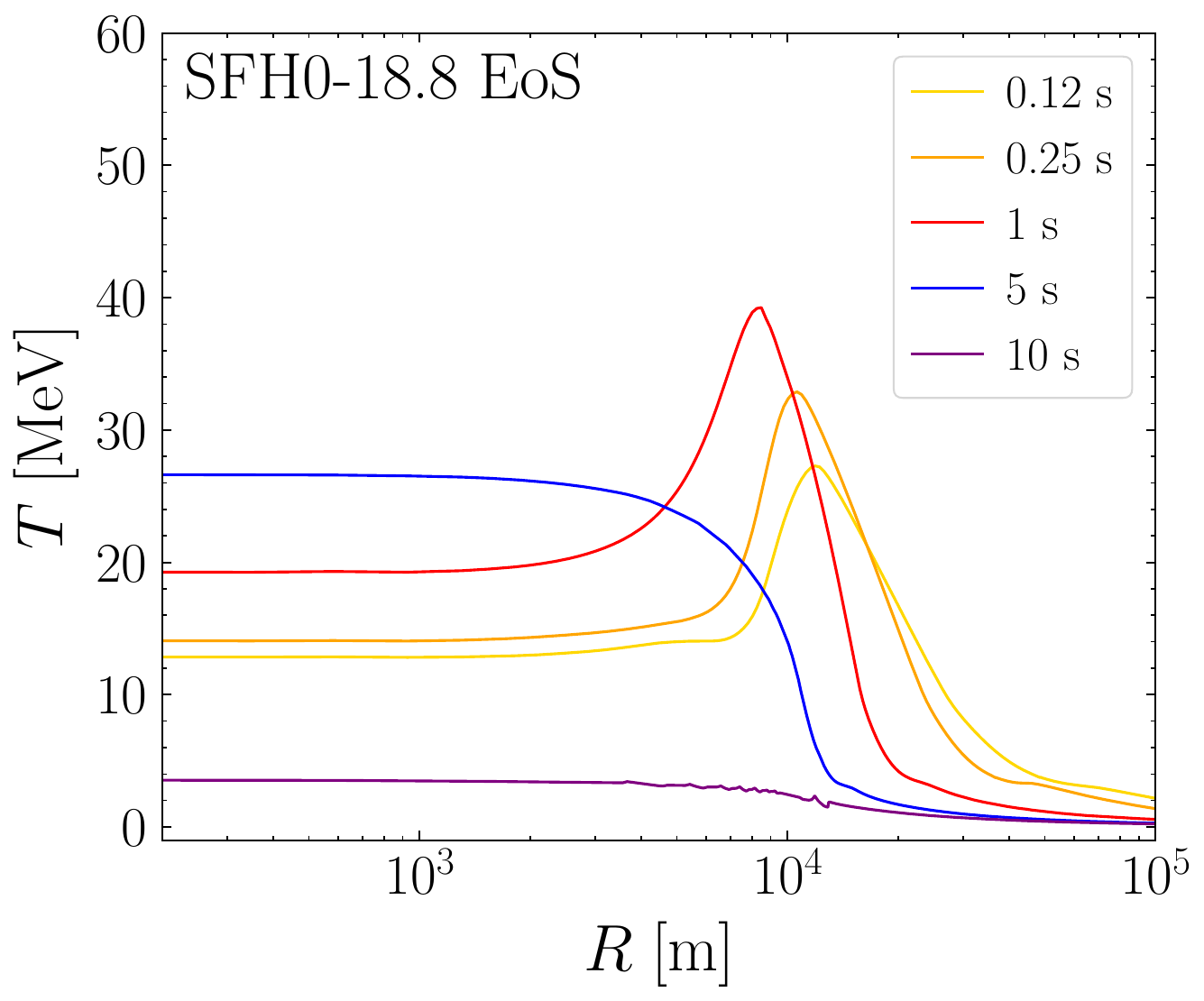}
    \includegraphics[width=0.3\columnwidth]{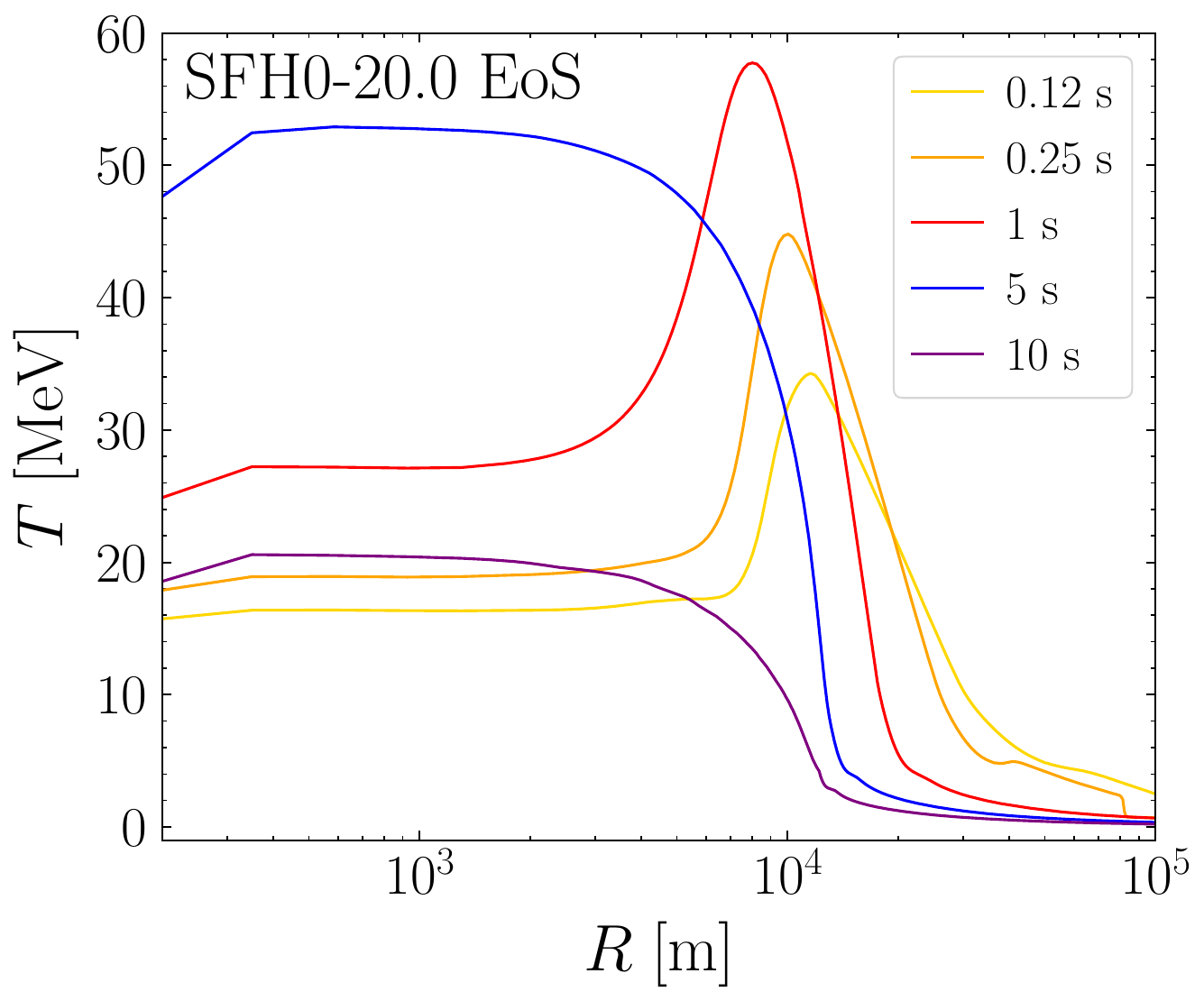}
    \caption{Temperature Profiles as a function of the radius $R$ within the PNS at different times for the three SN simulations used in this work.}
    \label{fig:TempProfs}
\end{figure}

\begin{figure}[h]
    \includegraphics[width=0.3\columnwidth]{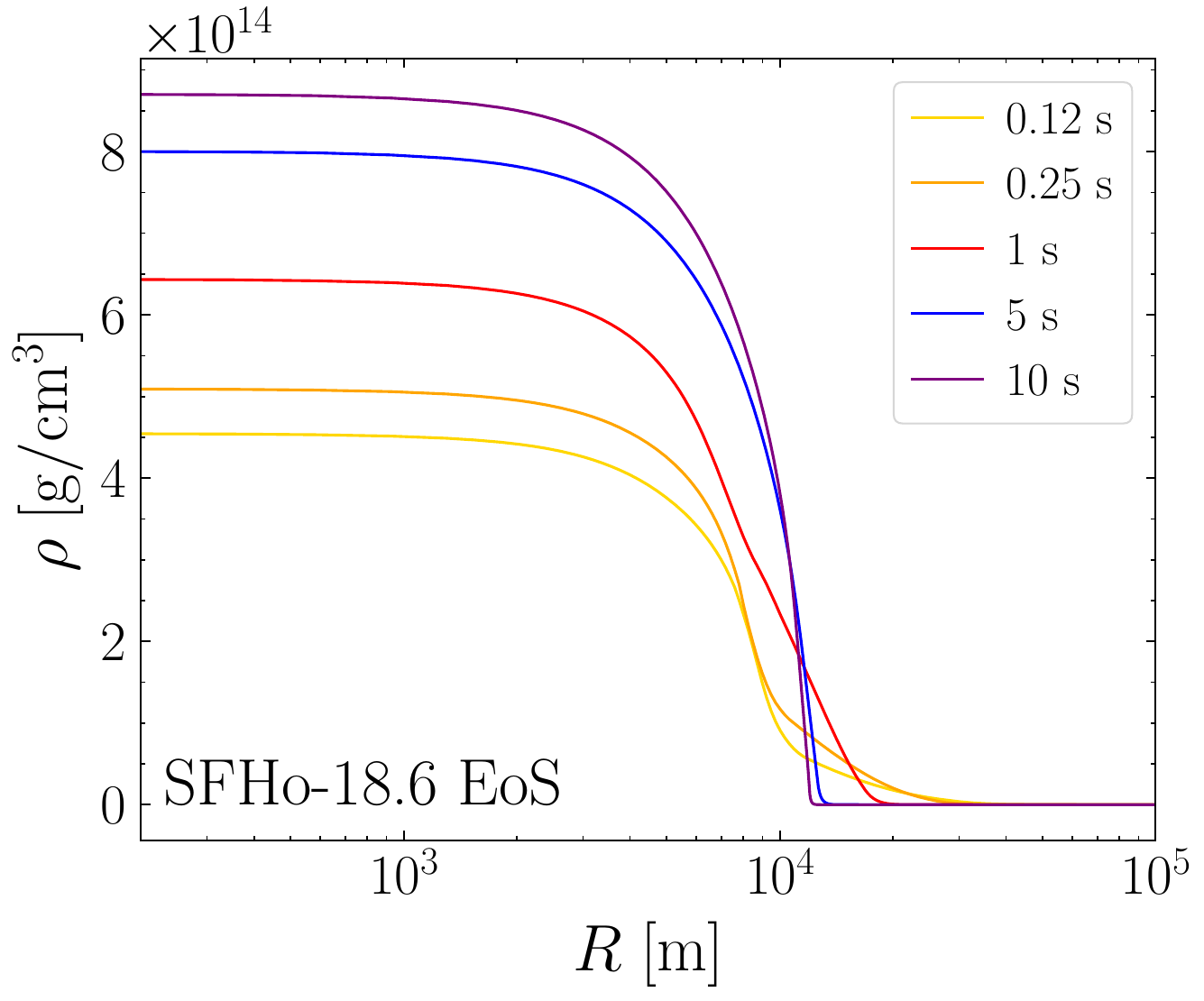}
    \includegraphics[width=0.3\columnwidth]{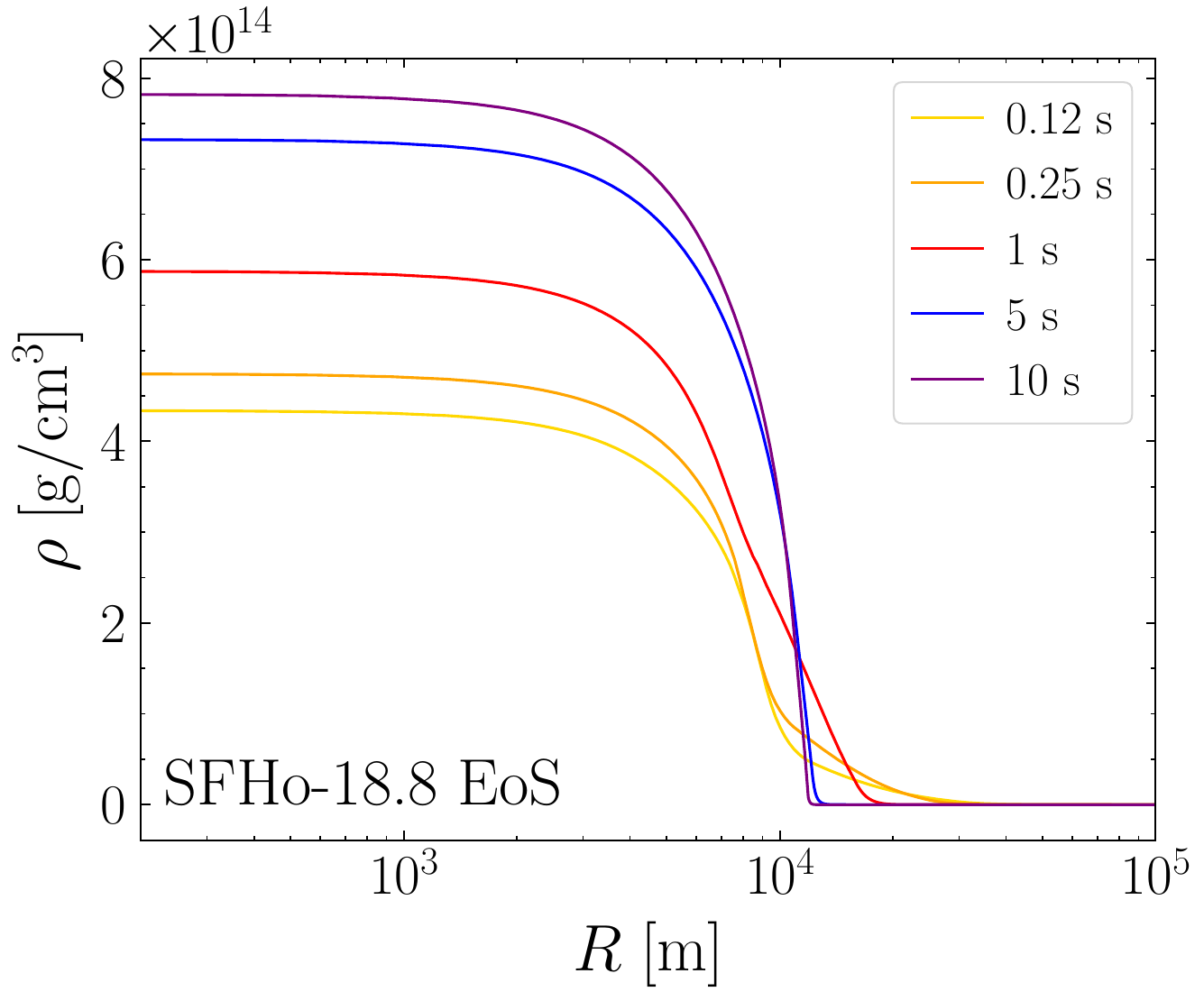}
    \includegraphics[width=0.3\columnwidth]{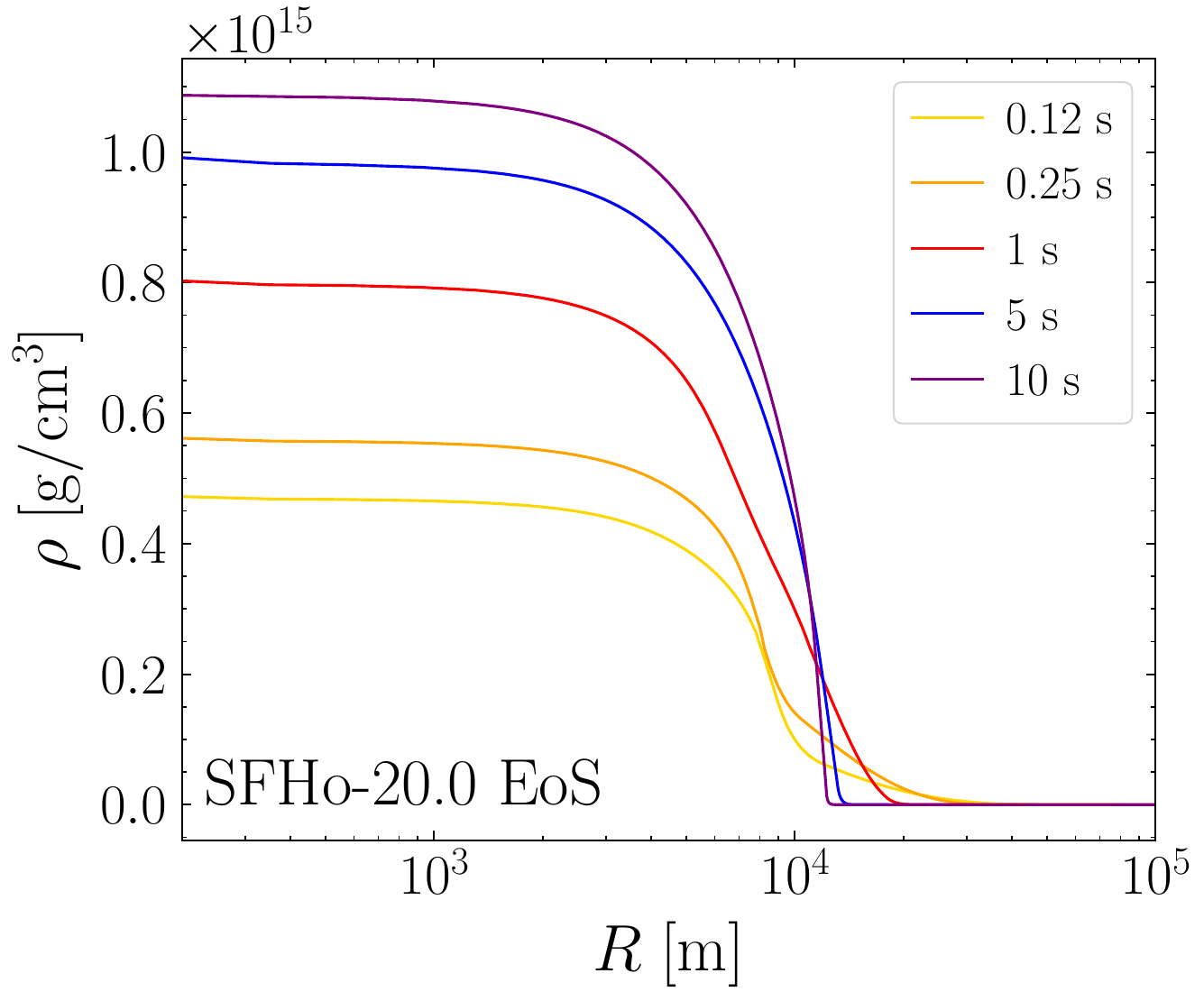}
    \caption{As in Fig.~\ref{fig:TempProfs} but for the nuclear density profiles.}
    \label{fig:DensProfs}
\end{figure}

\begin{figure}[h]
    \centering
    \includegraphics[width=0.45\columnwidth]{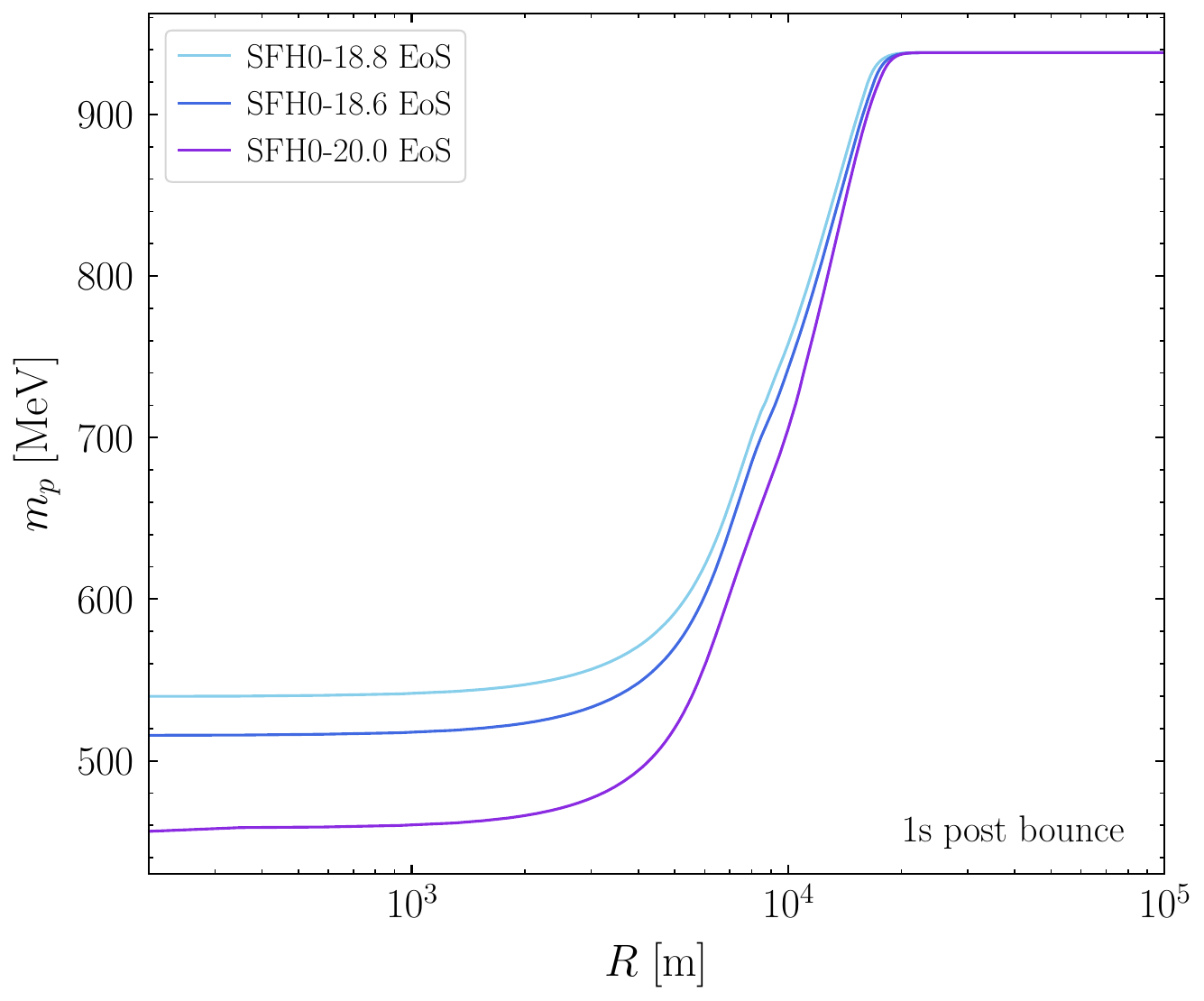}
    \includegraphics[width=0.45\columnwidth]{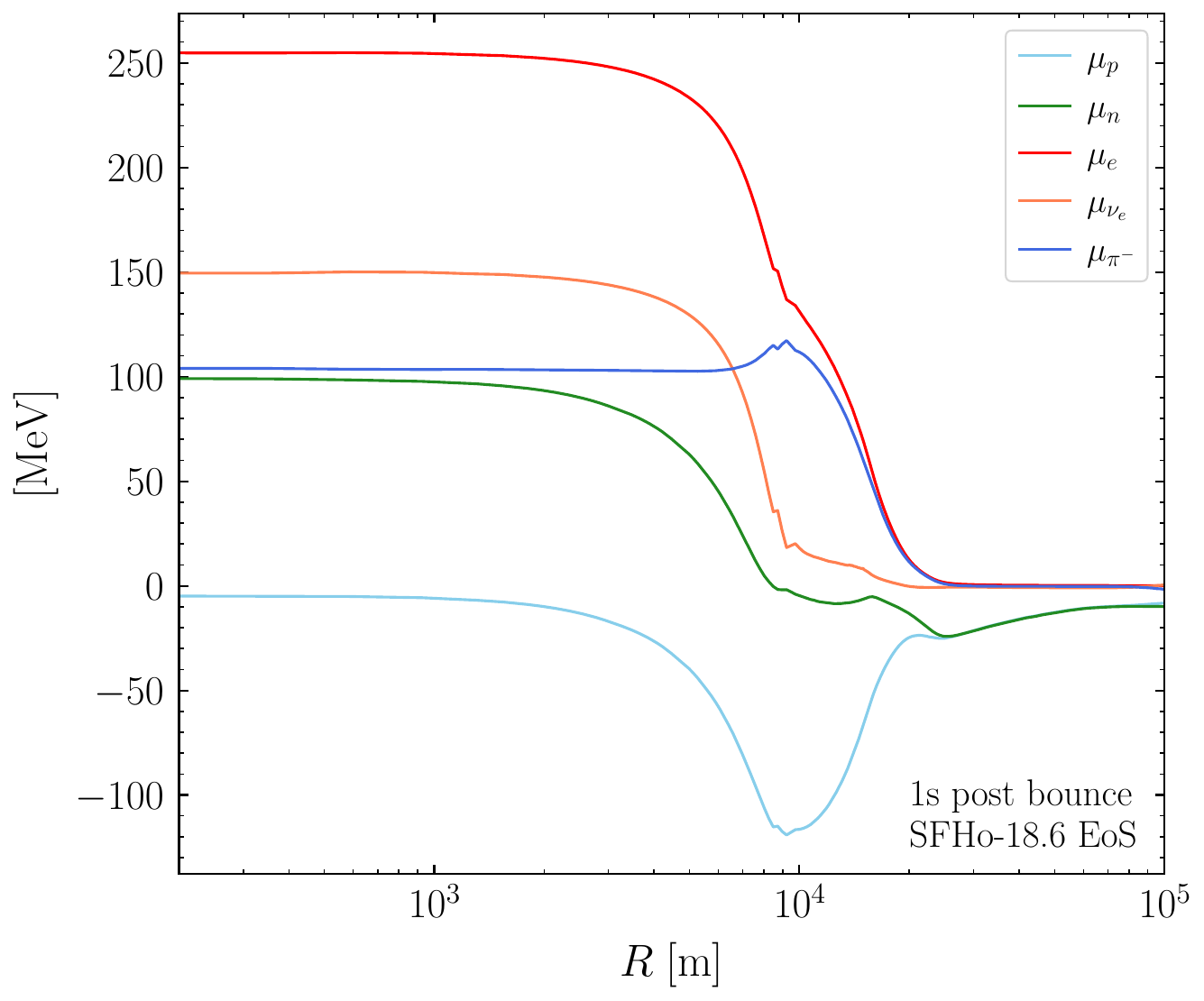}
    \caption{Profiles of the proton mass (left) and chemical potentials (right) at 1 s after the core collapse in the PNS.  In the left panel we illustrate the results across the three simulations considered, while in the right panel we show results for our fiducial simulation SFHo-18.6.  }
    \label{fig:pMassandChemPot}
\end{figure}

\begin{figure}[h]
    \centering
    \includegraphics[width=0.45\columnwidth]{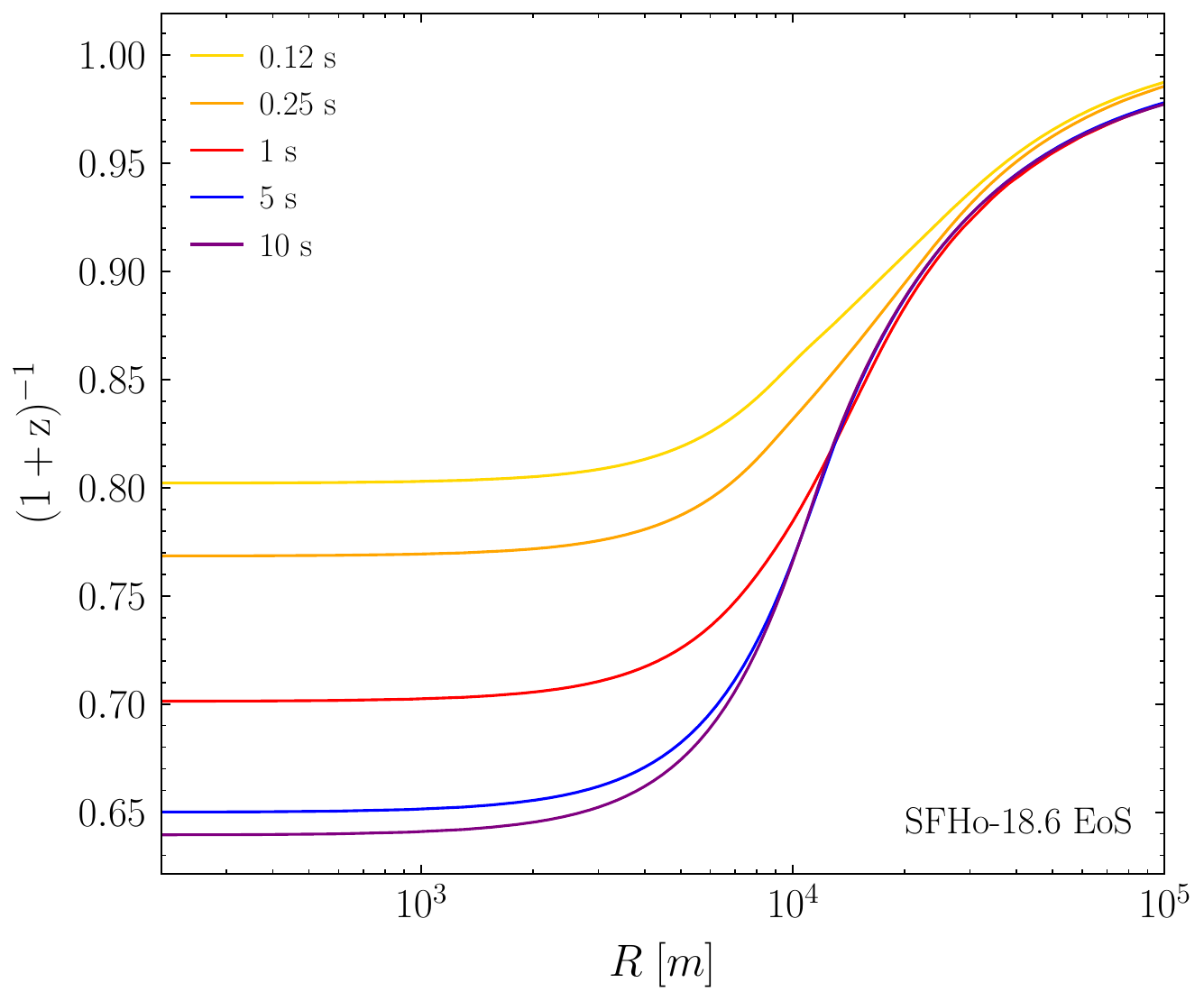}
    \caption{Profiles of $(1+z)^{-1}$, where $z$ is the gravitational redshift, obtained with our fiducial SN simulation SFH0-18.6 EoS, for different times after the SN explosion.}
    \label{fig:redshift}
\end{figure}

In this section we provide additional information related to the SN simulations used in this work from Ref.~\cite{Bollig:2020xdr}.
Figure~\ref{fig:TempProfs}, Fig.~\ref{fig:DensProfs} and Fig.~\ref{fig:pMassandChemPot} show the radial dependence of the temperature, density profiles, and proton mass along with chemical potentials (note that for nucleons we show the non-relativistic chemical potentials), respectively, across the three simulations considered in this work.  Figure~\ref{fig:redshift} shows the gravitational redshift $z$ (more precisely, we illustrate $(1+z)^{-1}$) due to the extreme densities reached in the SN core.   These quantities are used to compute the axion emissivity. 

In this work, we assume a thermal distribution of pions, with a chemical potential dictated by weak equilibrium, $\mu_{\pi^-} = \mu_e - \mu_{\nu_e}$. Note that the modeling of the pion abundance in dense nuclear matter is still an open question. The main uncertainties concern the presence and description of a Bose-Einstein condensation phase~\cite{Migdal:1990vm} as well as the effects of pion-nucleon interactions on the pion dispersion relation~\cite{Fore:2019wib, Carenza:2020cis, Fore:2023gwv}. Ultimately, dedicated SN simulations accounting for pions are needed to properly assess their impact on the SN thermodynamical evolution and axion luminosity. 

\section{Solar Maximum Mission Analysis and Comparison to GALAXIS / Fermi-LAT}

\begin{figure}[h!]
    \centering
    \includegraphics[width=0.8\columnwidth]{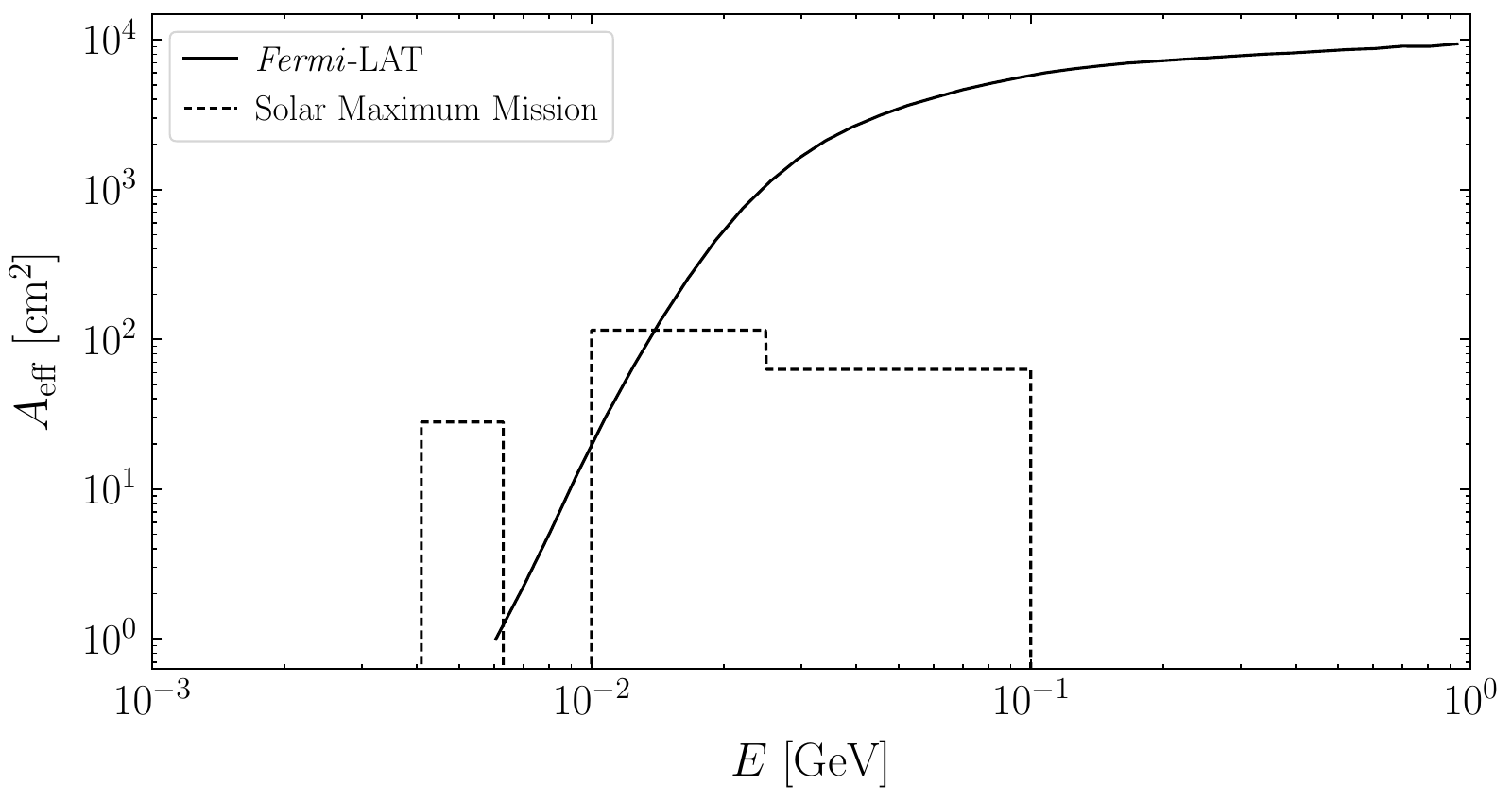}
    \caption{The effective area as a function of energy for both \textit{Fermi}-LAT (\texttt{P8R3\_TRANSIENT020\_V2} event class) and SMM. In our projections we assume that the proposed GALAXIS instrument has the Fermi-LAT on-axis instrument response but with $4 \pi$ angular coverage.}
    \label{fig:effective_area}
\end{figure}

We use the SN1987A data from the SMM provided in~\cite{Hoof:2022xbe} for the energy bins 4.1-6.4 MeV, 10-25 MeV, and 25-100 MeV, with time binned in $\Delta t = 2.048$ s intervals.
We model the data using a Poisson likelihood. For the background model contribution we use a linear ansatz of the form
\begin{equation}
    \mu^{\mathrm{bkg}}_{i, j} = b^{(0)}_{j} + b^{(1)}_{j} \frac{t_{i} - t_{\nu}}{\Delta t} \,,
\end{equation}
where $t_{i}$ corresponds to the $i^{\rm th}$ time bin and $j$ corresponds to the index over the three energy bins.  The six parameters $\{ b_j^{(0)},b_j^{(1)} \}_{j=1}^{3}$ are the background nuisance parameters. For the signal model, with mean prediction $\mu_{i,j}^{\rm sig}$, we take the axion emission spectrum and calculate the mean expected photon counts by convolving with the instrument response:
\begin{equation}
    \mu^{\mathrm{sig}}_{i, j} = \frac{A_{\mathrm{eff}, j}}{4 \pi d^2} \int_{E_{j}}^{E'_{j}} dE_{a} \int_{t_{i}}^{t'_{i}} dt \, P_{a \gamma}(g_{a \gamma \gamma}, m_{a}, E_{a}) \frac{d^2 N_{a}(g_{a\gamma\gamma},E_a,t)} {dt dE_{a}}  \,,
\end{equation}
where $P_{a \gamma}$ corresponds to the conversion probability of axions-to-photons, $d$ is the distance to the SN, $d^2N_a / dt / dE_a$ is the differential number of axions generated by the PNS per unit time per unit energy, $A_{ {\rm eff},j}$ is the SMM effective area in energy bin $j$, and $E_{j}$ $E'_{j}$ ($t_{i}$ $t'_{i}$) are the values of the bin edges for the $j^{\rm th}$ energy ($i^{\rm th}$ time) bin.  The values of $A_{ {\rm eff},j}$ are approximately $28, 115, 63$ cm$^2$ in energy bins $j = 1,2,3$, respectively.  In Fig.~\ref{fig:effective_area} we illustrate the SMM effective area and compare it to that of the Fermi-LAT. Note that we assume the proposed GALAXIS constellation has the same effective area as the Fermi-LAT but with full-sky angular coverage.  In addition to gaining in effective area, the Fermi-LAT also has significantly improved background rate with respect to SMM. For Fermi-LAT (and thus also for GALAXIS) we project zero background events within the $\sim$10 s of the SN event, while for SMM the number of background events was around 50 events per 10 s.  

Note, also, that while we assume that the GALAXIS network will have $4 \pi$ angular coverage, it is possible that the network could still have a high chance of detecting the next Galactic SN with slightly less angular coverage, considering that the next SN will likely lie in the Galactic plane. We leave a technical optimization of the sky coverage for future work.

We analyze the data using the joint signal plus background model, fixing the start of the SN to the observed time from the neutrino burst. We include data from $\sim$150 s before the event until $\sim$220 s afterwards.  At fixed $m_a$ we compute the profile likelihood as a function of the signal strength parameter $g_{a \gamma\gamma}$ profiling over the background nuisance parameters. Note that for consistency, to account for downward fluctuations, we also allow for negative signal strengths; we then use Wilks' theorem, relying on the observation that the numbers of counts per bin are typically above 10, to set the 95\% upper limits and compute the discovery test statistics (see, {\it e.g.},~\cite{Safdi:2022xkm}).

\begin{figure}[h]
    \centering
    \includegraphics[width=0.8\columnwidth]{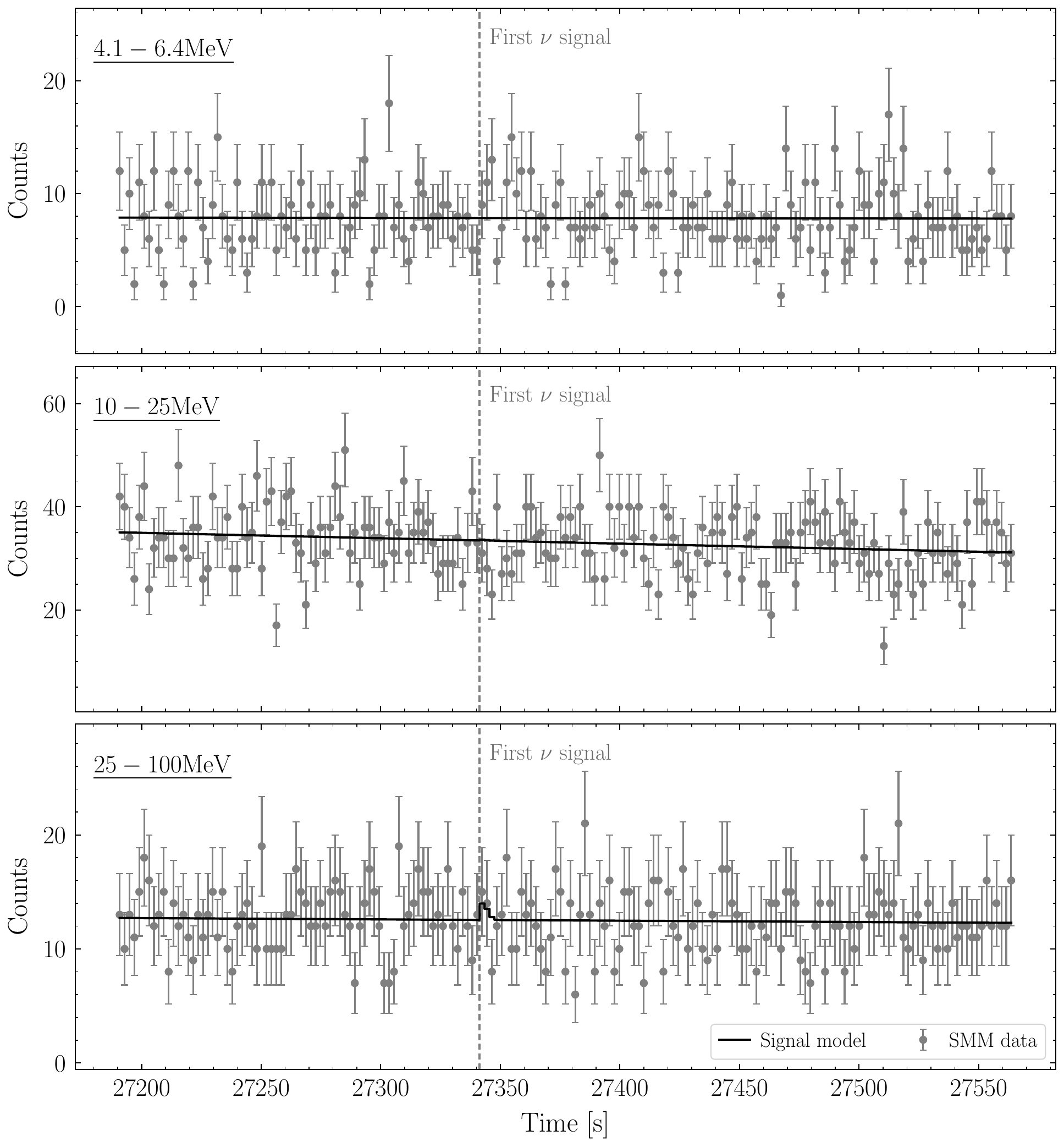}
    \caption{The SMM data provided in \cite{Hoof:2022xbe} in the three energy bins within the time-frame of SN1987A. We illustrate the best-fit signal plus background model at $m_a = 0$ accounting for axion-to-photon conversion in the stellar magnetic field of the progenitor star, assuming $B_0 = 1$ kG, and the Galactic magnetic field; in this case the best-fit coupling strength is 
    % $g_{a \gamma \gamma} \approx 5.6 \times 10^{-12}$ GeV$^{-1}$ 
    \new{$g_{a \gamma \gamma} \approx 2.4 \times 10^{-12}$ GeV$^{-1}$} and the discovery test statistic is 
    % ${\rm TS} \approx 0.3$
    \new{${\rm TS} \approx 0.32$}.}
    \label{fig:smm_data}
\end{figure}

The data are illustrated in Fig.~\ref{fig:smm_data} along with the best-fit signal plus background model for the axion-like particle scenario with $m_a = 0$ eV accounting for conversion in the stellar magnetic field with $B_0 = 1$ kG and the Galactic magnetic field; the best-fit coupling is 
\new{$g_{a \gamma \gamma} = 2.4 \times 10^{-12}$ GeV$^{-1}$}
% g_{a\gamma\gamma} = 5.6 \times 10^{-12}$ GeV$^{-1}$
in that case, with discovery test statistic
\new{${\rm TS} \approx 0.32$}.
% ${\rm TS} \approx 0.3$.

\section{Analysis Variation}

\begin{figure}[h]
    \centering    \includegraphics[width=0.8\columnwidth]{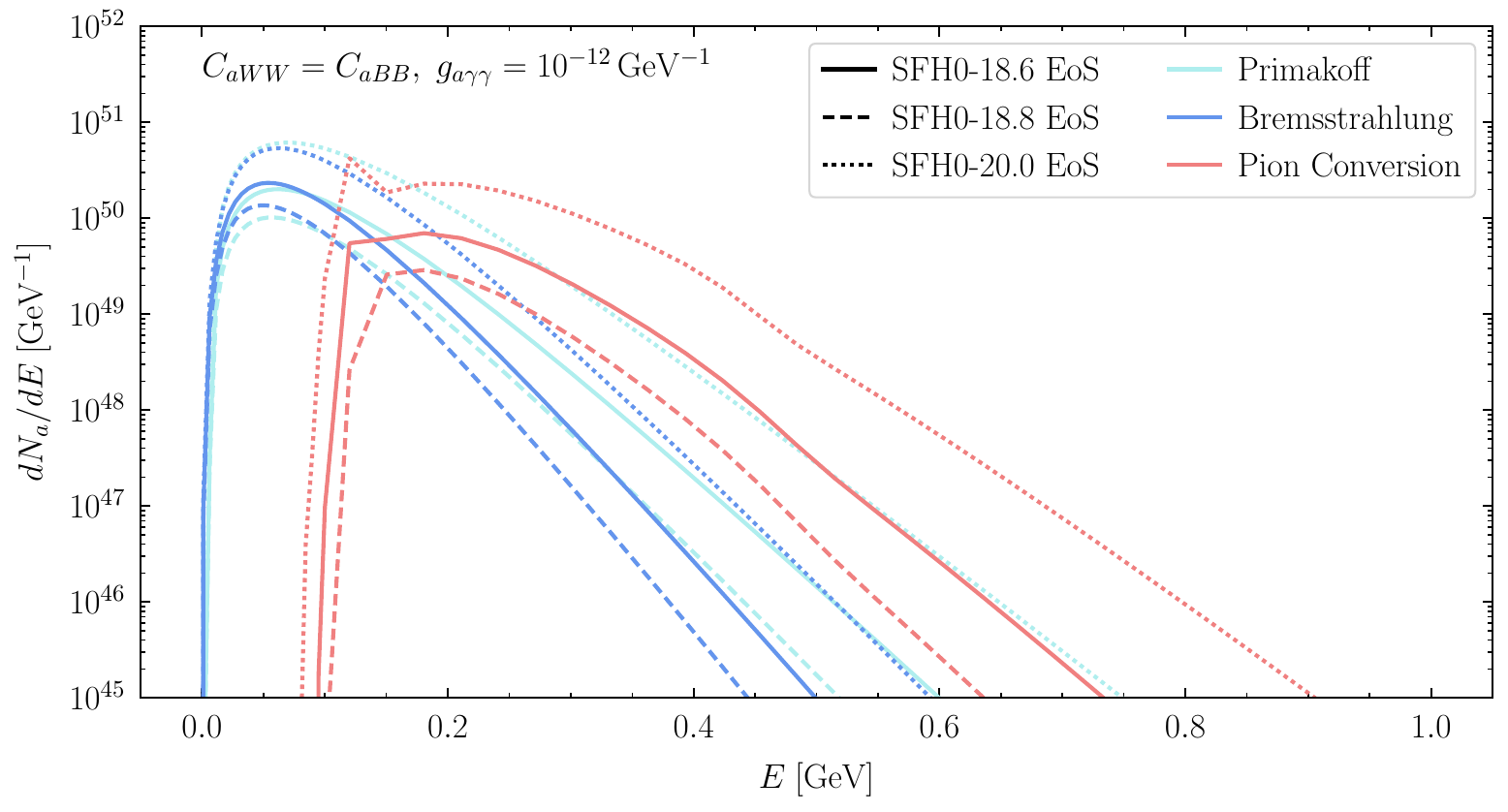}
    \caption{The differential axion spectra, integrated over the first 10 s after the core collapse, obtained from different processes, using the three SN simulations discussed in Sec.~\ref{sec:SNSim}.  Note that we assume $C_{aWW} = C_{aBB}$ and take $g_{a\gamma\gamma} = 10^{-12}$ GeV$^{-1}$ for definiteness in this figure. }
    \label{fig:EnergyDistr}
\end{figure}

\begin{figure}[h]
    \centering
    \includegraphics[width=0.8\columnwidth]{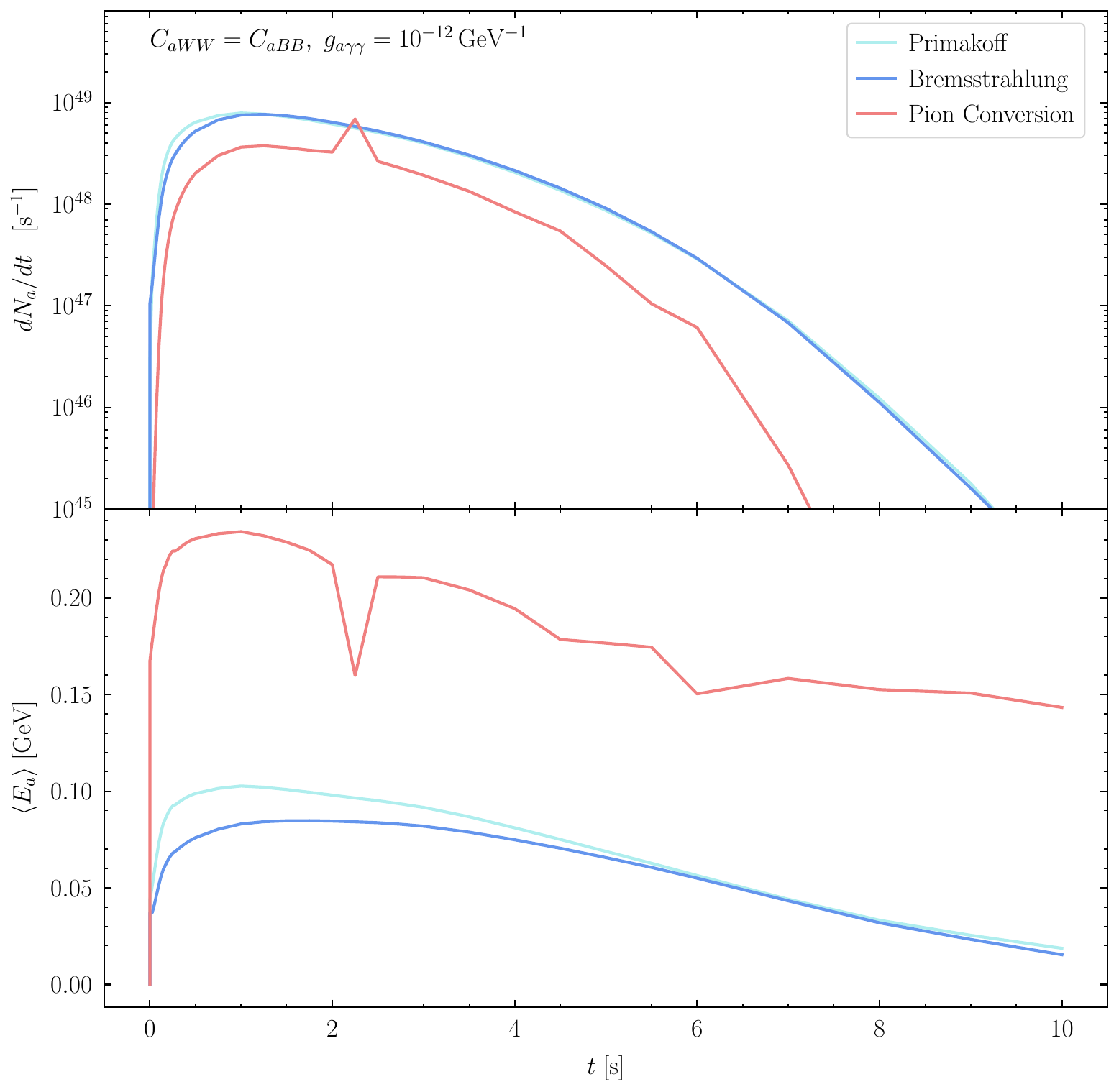}
    \caption{(Top panel) The time dependence of the axion luminosity, integrated over energy for the different production mechanisms discussed.  We illustrate these results for our fiducial SN simulation (SFH0-18.6 EoS).  (Bottom panel) The average energy of emitted axions broken up by production mechanism.  Pionic processes result in more energetic axions.  The sharp feature in the pion emission around $t \sim 2$ s is due to an incorrect, but also insignificant, description of the pion condensate in our formalism (see text for details).} %
    \label{fig:TimeDistr}
\end{figure}

In this section we summarize a number of analysis variations that are aimed to assess the robustness of the results presented in the main Letter.
First, in Fig.~\ref{fig:EnergyDistr} we show comparisons of the differential axion spectra, integrated over time, for Primakoff, bremsstrahlung and pion conversion, obtained with the three different simulations discussed in Sec.~\ref{sec:SNSim}.  Pion conversion dominates in all cases, though there is roughly an order of magnitude spread in the predicted spectra across the simulations.  In Fig.~\ref{fig:TimeDistr} (top panel) we show the mean differential number of axions generated in the PNS, integrated over energy, as a function of time for our fiducial SN simulation. We again separately break down the contributions from Primakoff, bremsstrahlung and pion conversion.  The bottom panel of Fig.~\ref{fig:TimeDistr} illustrates the mean energy $\langle E_a \rangle$ of axions emitted from the SN for each process. The pion conversion processes emit more energetic axions than Primakoff and bremsstrahlung. 

\begin{figure}
    \centering
    \includegraphics[width=0.8\columnwidth]{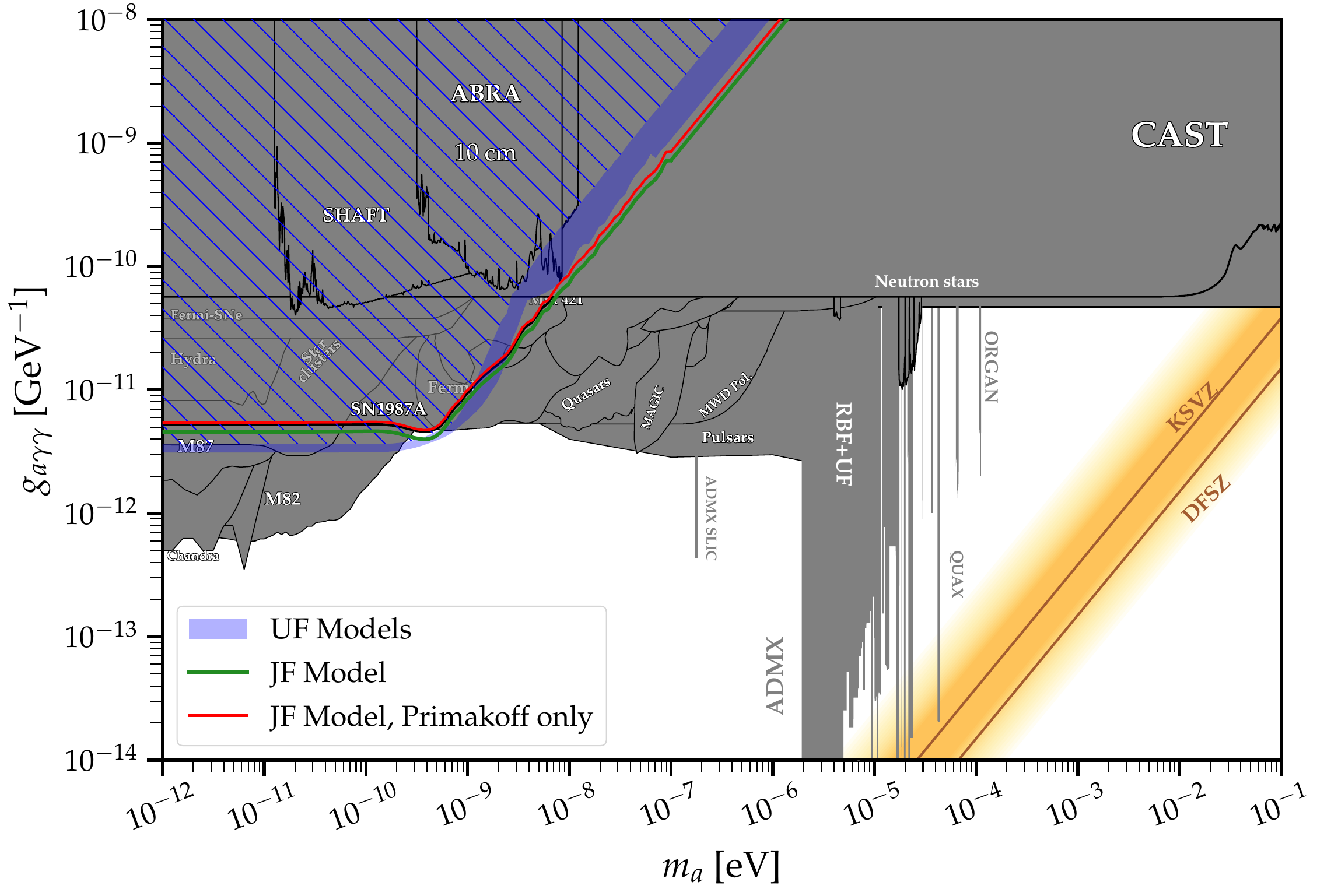}
    \caption{
    As in Fig.~\ref{fig:SN_limit_projs} for the low-mass SN1987A limits accounting for axion-to-photon conversion on the Galactic magnetic fields, but varying over magnetic fields models. Note that for the 8 UF models we shade the range of upper limits found across those models; our fiducial limit is, at each $m_a$, taken to be the weakest across that ensemble. 
    }
    \label{fig:LimitsGalactic}
\end{figure}

\begin{figure}
    \centering
    \includegraphics[width=0.8\columnwidth]{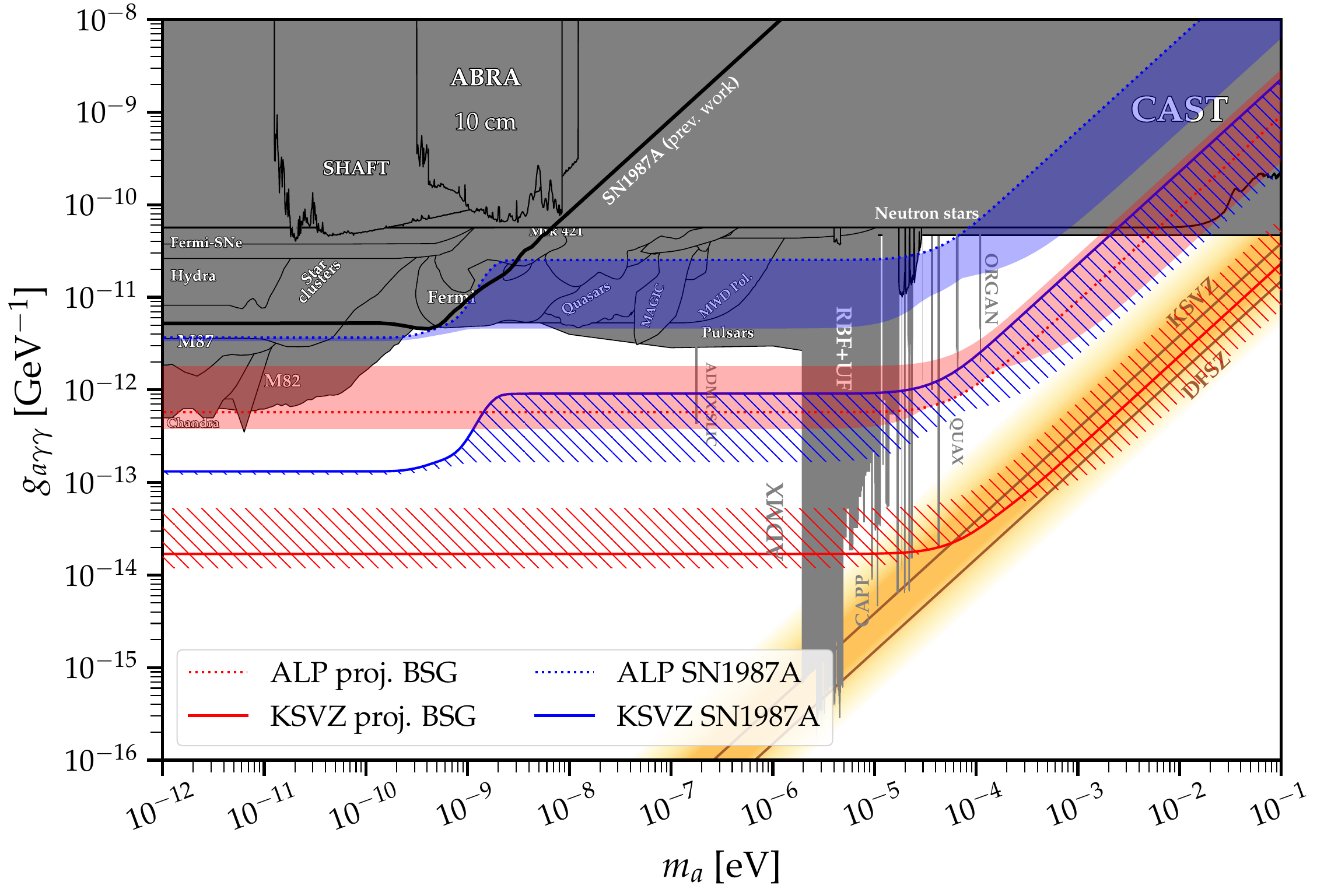}
    \caption{
     As in Fig.~\ref{fig:SN_limit_projs} but only showing the results for conversion on the stellar magnetic field and varying the magnetic field strength at the surface of the BSG between $B_0 = 100$ G and $B_0 = 10$ kG. 
    }
    \label{fig:Bvariation}
\end{figure}

\begin{figure}
    \centering
    \includegraphics[width=0.8\columnwidth]{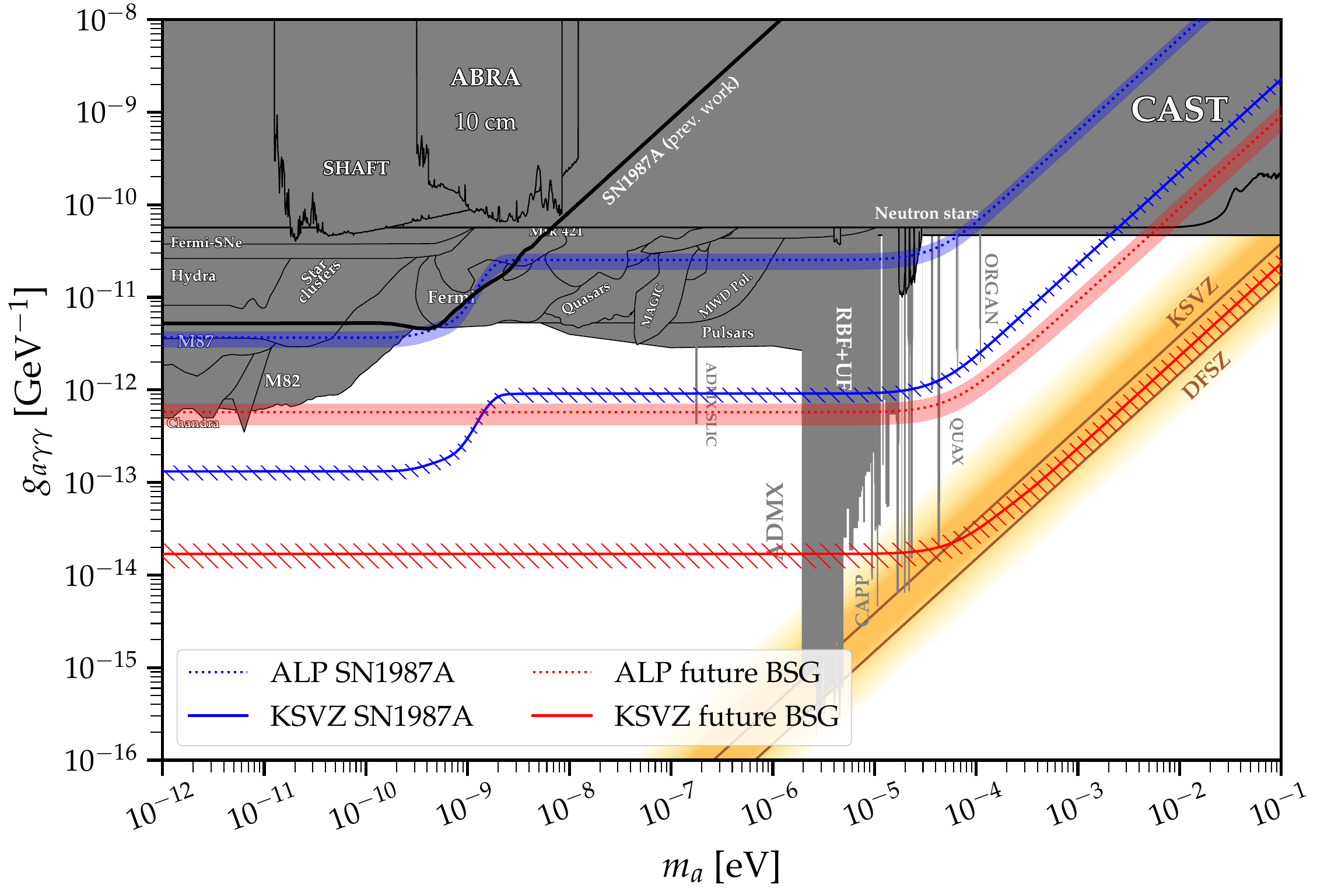}
    \caption{ As in Fig.~\ref{fig:SN_limit_projs} but varying between the three SN simulations discussed in this work, with all other parameters as in Fig.~\ref{fig:SN_limit_projs}. The stronger limits arise from more massive PNSs.  The curves within the shaded regions show the fiducial results used in Fig.~\ref{fig:SN_limit_projs}.
    }
    \label{fig:SimVariation}
\end{figure}

In Fig.~\ref{fig:LimitsGalactic} we compare the SN1987A limits obtained using different models for the Galactic magnetic field (see Sec.~\ref{sec:gal} for a discussion). 
These limits are obtained with the simulation SFHo-18.6, assuming a BSG progenitor star with a radius of 45 $R_\odot$ as in our fiducial limits.
On the other hand, in Fig.~\ref{fig:Bvariation} we consider our fiducial SN1987A scenario but vary the surface magnetic field $B_0$ from 100 G to 10 kG to, broadly, encapsulate the uncertainty in the magnetic field for a typical BSG. (Note that the 100 G assumption leads to the weakest limits, with 10 kG giving the strongest results.)  We also show, as in Fig.~\ref{fig:SN_limit_projs}, the projections for a next BSG Galactic SN. 
The analogous results obtained with the three different SN simulations, assuming a BSG progenitor star with a radius of 45 $R_\odot$ and a magnetic field at the surface of the star of 1 kG, are shown in Fig.~\ref{fig:SimVariation}.

\begin{figure}
    \centering
    \includegraphics[width=0.8\columnwidth]{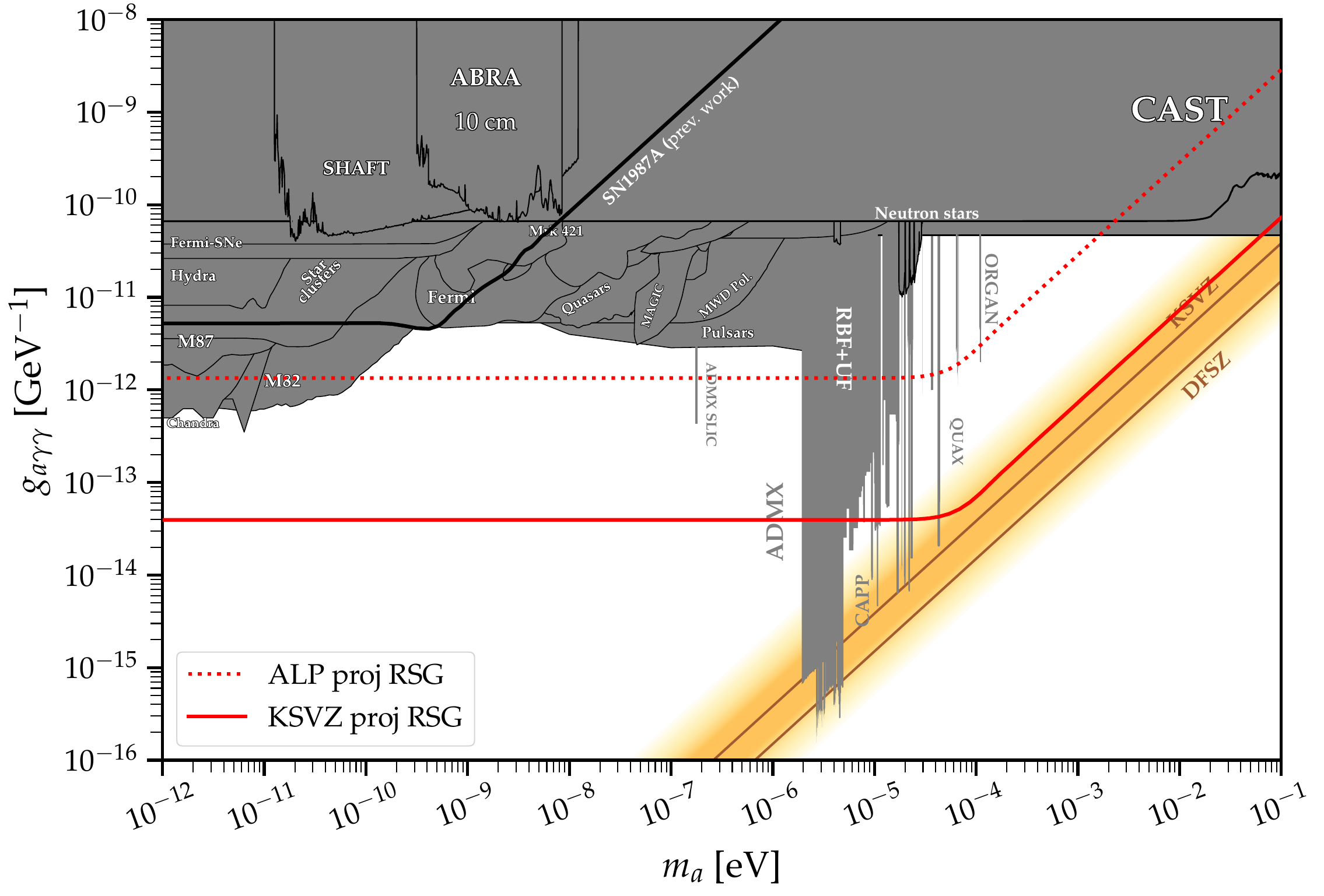}
    \caption{
    As in Fig.~\ref{fig:SN_limit_projs} but only showing projections for the next Galactic SN and assuming a RSG progenitor at 10 kpc from Earth instead of a BSG progenitor.  The RSG projections are slightly less strong than those for the BSG, though we stress that the RSG results are subject to large uncertainties on the magnetic field modeling.}
    \label{fig:SimVariation_RSG}
\end{figure}

In Fig.~\ref{fig:SimVariation_RSG} we project the upper limits under the null hypothesis, as in Fig.~\ref{fig:SN_limit_projs}, but assuming a RSG progenitor. As discussed more in Sec.~\ref{sec:RSG}, RSGs are more likely core-collapse SN progenitors than BSGs, but their magnetic field distributions are more uncertain as the relevant fields arise dynamically from the conducting fluid dynamics in the outer layers of the star. 

\begin{figure}
    \centering
    \includegraphics[width=0.8\columnwidth]{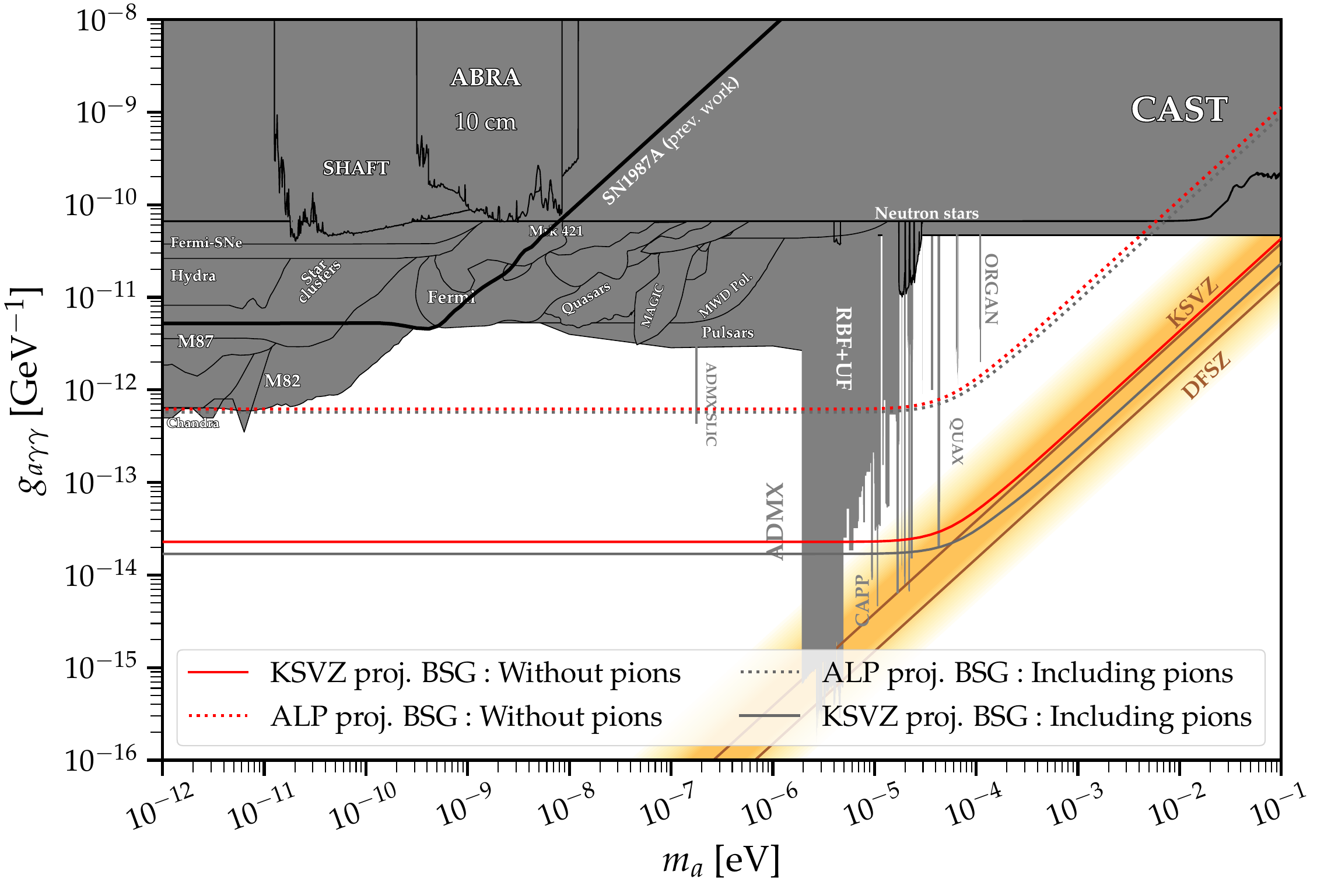}
    \caption{As in Fig.~\ref{fig:SN_limit_projs} for future SN projections with GALAXIS including (as we do in our fiducial calculations) or removing the axion production from axion-pion conversion.}
    \label{fig:no_pion}
\end{figure}

Given the uncertainties in the pion-conversion processes, it is instructive to illustrate projected upper limits with GALAXIS for the next Galactic SN assuming no pion-induced emission processes.  As seen in Fig.~\ref{fig:no_pion}, including pions does not strongly affect the low-mass sensitivity and has a slightly more pronounced affect at higher masses, since the pions give rise to higher-energy axions relative to {\it e.g.} bremsstrahlung processes.  Note that pion emission is not relevant for SN1987A given the limited energy range of the SMM instrumentation. 

\begin{figure}
    \centering
    \includegraphics[width=0.49\columnwidth]{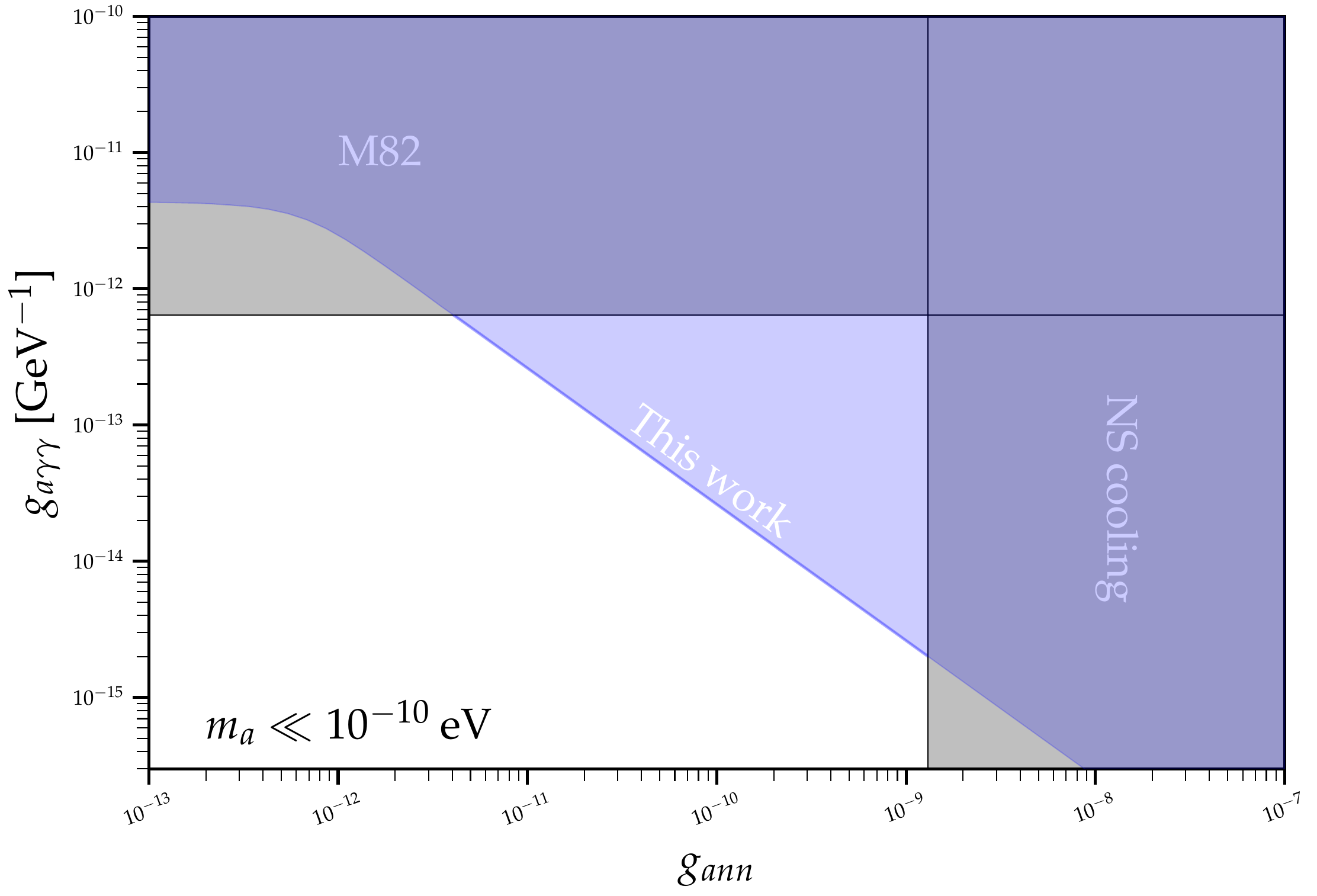}
    \includegraphics[width=0.49\columnwidth]{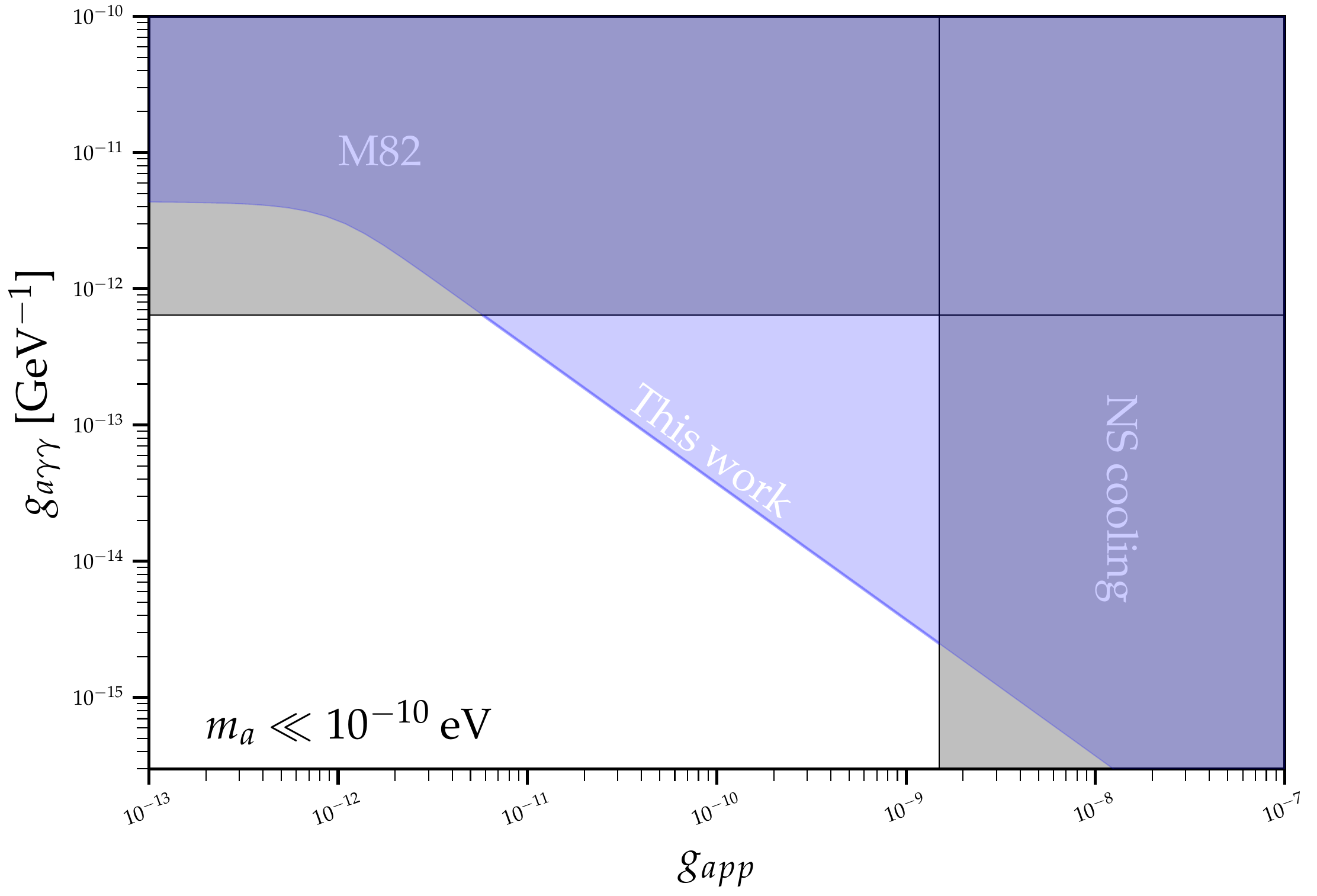}
    \caption{Upper limits computed in this work on the EFT parameter space $g_{a\gamma\gamma}-g_{ann}$, with $g_{app} = 0$ (left panel), and $g_{a\gamma\gamma}-g_{app}$, with $g_{ann} = 0$ (right panel), from the non-observation of gamma-rays from SN1987A. We account for axion-to-photon conversion in the Galactic magnetic field and assume $m_a \ll 10^{-10}$ eV such that the axion mass does not play a role in the axion-photon conversion process. Our new upper limits exclude previously allowed regions of the illustrated parameter space.}
    \label{Fig:axion_plane}
\end{figure}

Lastly, it is useful to rephrase our limits from SN1987A in terms of limits in the space of $g_{a\gamma\gamma}$, $g_{ann}$, and $g_{app}$ without assuming any relations between these EFT parameters. However, to simplify the parameter space we assume $m_a \ll 10^{-10}$ eV so that we may neglect the axion mass when computing the Galactic axion-to-photon conversion probabilities. The limits illustrated in Fig.~\ref{fig:SN_limit_projs} may then be expanded into limits on $g_{a\gamma\gamma}$-$g_{ann}$ (assuming $g_{app} = 0$) and $g_{a\gamma\gamma}$-$g_{app}$ (assuming $g_{ann} = 0$); see Fig.~\ref{Fig:axion_plane}.  These new upper limits computed in this work surpass those on the axion-photon coupling only ({\it e.g.},~\cite{Reynes:2021bpe,Ning:2024eky}) and the nucleon couplings only ({\it e.g.},~\cite{Buschmann:2021juv}) for some of the parameter space.

\section{Neutron star merger estimates}

In the main Letter we focus on axion-induced gamma-ray signals from PNSs formed after core-collapse SN. On the other hand, it is also interesting to consider similar signals created from the hot PNSs formed after NS mergers. A key difference between SN and NS mergers for the signal of interest is that for SN the axions must escape the much larger surrounding star before converting to gamma-rays, while for NS mergers axions may immediately begin to convert to gamma-rays after exiting the PNS. 

Modeling the axion-induced gamma-ray signal from NS mergers would require dedicated simulations to accurately model the PNS interior and the magnetic field distributions. Such simulations, incorporating axion emission, would be an interesting direction for future work. Here, on the other hand, we apply more basic approximations in order to show that NS mergers are potentially promising targets for axion searches in gamma-rays.

The first approximation we make in this work is that the integrated (over time) differential (in terms of energy) flux of axions produced during a NS merger matches that in our default SN simulation.  This approximation is almost certainly not correct in detail, but we may roughly compare the predicted axion luminosities between SN and NS-merger remnants through the following argument.  The axion luminosities from nuclear matter in the degenerate limit roughly scale as $T^6$ (see SM Secs.~\ref{sec:pion_axion} and~\ref{sec:nucelon_axion}).  A NS remnant may have a maximum temperature $\sim$40-60 MeV for a time of around 1 s (see, {\it e.g.}, Fig.~\ref{fig:TempProfs}).  On the other hand, the remnants formed immediently before and during NS mergers can have much higher temperatures over shorter periods of time.

NS mergers may result in black holes or in stable NS remnants. In the case of black hole formation, the hypermassive remnant still exists for $\sim$20 ms after merger, until angular momentum loss from gravitational wave radiation causes the remnant to collapse (see, {\it e.g.},~\cite{Kiuchi:2022nin} and references therein). During the initial $\sim$10 ms the temperature can potentially surpass 100 MeV regardless of whether the remnant is stable or not~\cite{Hanauske:2019qgs,Dietrich:2019shr,Zappa:2022rpd,Kiuchi:2022nin,Radice:2023zlw}.  We may then roughly estimate the axion luminosity of a NS merger, regardless of the final state, relative to that from a SN by $(100 \, \, {\rm MeV} / 50 \, \, {\rm MeV})^6 \times (10 \, \, {\rm ms} / 1 \, \, {\rm s}) \sim 0.6$. That is, even though remnants from NS mergers (for example, those collapsing to black holes) may also survive for fractions of a second, the high temperatures reached in these dense systems can potentially lead to axion luminosities comparable to those found in the PNSs formed after SN.  Of course, the estimate above is extremely rough, but it justifies -- on a first pass -- using a SN simulation to model the remnant formed after the NS merger. A more accurate projection would used dedicated NS merger simulations, which we leave for future work. 

\begin{figure}
    \centering
    \includegraphics[width=0.8\columnwidth]{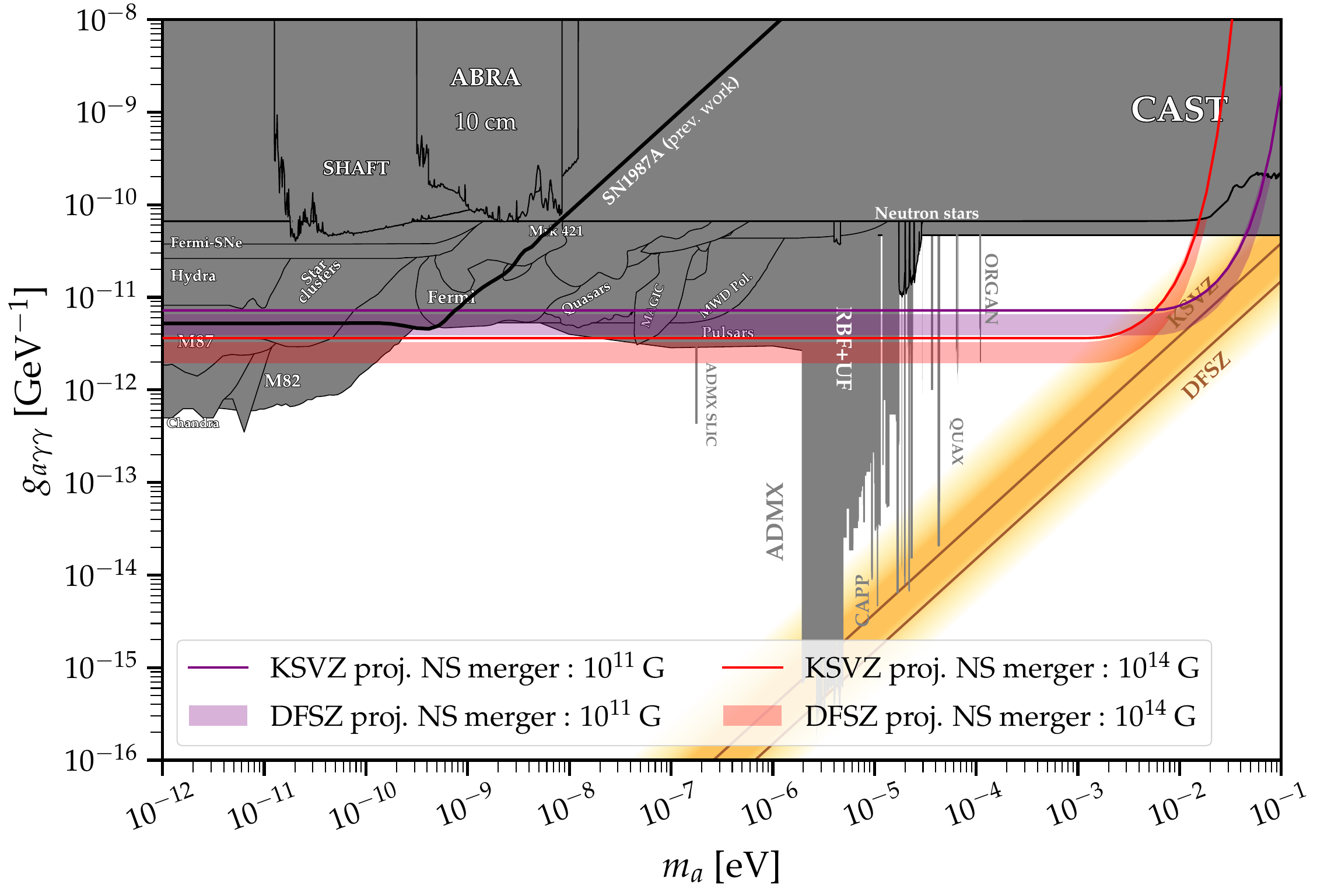}
    \caption{Projected 95\% upper limits under the null hypothesis on $g_{a\gamma\gamma}$ from an observation of Fermi / GALAXIS towards a NS-NS merger at a distance of 30 Mpc. As a rough estimate, we use the SFHo-20.0 SN simulation to describe the remnant from the merger. We also model the magnetic field of the remnant as a dipole with surface field strengths as indicated (we consider two possibilities). These approximations are rough but motivate future work that more accurately computes the axion-induced signal from NS mergers. }
    \label{fig:NS_merger}
\end{figure}

As a very rough approximation to a NS merger, we thus use our SN simulation SFHo-20.0, which results in the most massive remnant of our SN simulation suite. (Again, we caution that this is not an accurate simulations of a NS merger.)  In Fig.~\ref{fig:NS_merger} we project the sensitivity of Fermi-LAT (or our proposed GALAXIS) to the axion-signal from a NS merger using the SFHo-20.0 simulation to describe the remnant evolution.\footnote{A NS-NS merger may give higher energy axions than those found in the SFHo-20.0 simulation, which would extend the reach to higher mass axions versus what is shown in Fig.~\ref{fig:NS_merger}.} We assume a distance of 30 Mpc to the NS merger.  This distance estimate is motivated by {\it e.g.} the 2017 NS merger resulting in the gravitational wave signal GW170817
detected by LIGO~\cite{LIGOScientific:2018cki}; this event originated $\sim$40 Mpc from Earth.  Events at $\sim$30 Mpc may potentially be expected around once per 10 years~\cite{LIGOScientific:2018mvr}. (Note that the upper limits on $g_{a\gamma\gamma}$ scale with the distance to the SN as $g_{a\gamma\gamma} \propto \sqrt{d}$, so the difference between 30 Mpc and {\it e.g.} 50 Mpc is relatively minor.)

In Fig.~\ref{fig:NS_merger} we model the magnetic field of the remnant as a centered dipole with surface field strength (at a radius $r_0 = 10$ km) of $B_0 = 10^{11}$ G and $B_0 = 10^{14}$ G. The first choice of field is motivated by the typical magnetic field strength of older NSs, while the second accounts for the expected amplification in NS mergers. In fact, magnetic fields can be amplified during the NS merger to well above $\sim$$10^{15}$ G~\cite{Combi:2023yav}.  Moreover, the magnetic fields generated during this amplification are not regular but turbulent, which would increase the conversion probability for high-mass axions by giving the B-field non-trivial spatial dependence.  Still, for the purpose of the simple estimates in this work we adopt the dipole field configurations.  Then, we compute the axion-to-photon conversion probabilities using the approximate analytic relations in~\cite{Fortin:2021sst}.  We illustrate both KSVZ and DFSZ sensitivities in Fig.~\ref{fig:NS_merger}; to conclude, we note that due to the compact nature of these remnants the projected upper limits become tantalizingly close to probing the QCD axion for masses $m_a \sim 10^{-3} - 10^{-2}$ eV, motivating incorporating axions into dedicated simulations.

\end{document}